\documentclass[aps,prl,twocolumn,amsmath,amssymb,floatfix,longbibliography]{revtex4-2}
	
\usepackage{amsmath}
\usepackage{txfonts}
\usepackage{microtype}
\usepackage{graphicx}
\usepackage{color}
\usepackage{ulem}
\usepackage{bm}
\usepackage[english]{babel}

\begin{document}

\title{Reconciling conflicting approaches for the tunneling time delay in strong field ionization}

\author{M. Klaiber}
\author{Q. Z. Lv}
\author{S. Sukiasyan}
	\author{D. Bakucz Can\'{a}rio}
\author{K. Z. Hatsagortsyan}\email{k.hatsagortsyan@mpi-hd.mpg.de}
\author{C. H. Keitel}
\affiliation{Max-Planck-Institut f\"{u}r Kernphysik, Saupfercheckweg 1,
	69117 Heidelberg, Germany}

\date{\today}

\begin{abstract}

 Several recent attoclock experiments have investigated the fundamental question of a quantum mechanically induced time delay in tunneling ionization via extremely precise photoelectron momentum spectroscopy. The interpretations of those attoclock experimental results were controversially discussed, because the entanglement of the laser and Coulomb field did not allow for theoretical treatments without undisputed approximations. The method of semiclassical propagation matched with the tunneled wavefunction, the quasistatic Wigner theory, the analytical R-matrix theory, the backpropagation method, and the under-the-barrier recollision theory are the leading conceptual approaches put forward to treat this problem, however, with seemingly conflicting conclusions on the existence of a tunneling time delay. To resolve the contradicting conclusions of the different approaches, we consider a very simple tunneling scenario which is not plagued with complications stemming from the Coulomb potential of the atomic core, avoids consequent controversial approximations and, therefore, allows us to unequivocally identify the origin of the tunneling time delay.

\end{abstract}

\date{\today}

\maketitle

Time delay in tunneling is a fascinating fundamental quantum problem, most recently measured in an experiment with cold atoms \cite{Ramos_2020}. In particular, there has been intense and often controversial discussion about  time delay in strong field tunneling ionization  \cite{Eckle_2008a,Eckle_2008b,Pfeiffer_2012,Landsman_2014o,Camus_2017,Sainadh_2019,Czirjak_2000,Lein_2011,
Orlando_2014,Orlando_2014PRA,Landsman_2014,Torlina_2015,Landsman_2015,Teeny_2016a,Teeny_2016b,Yakaboylu_2013,Klaiber_2018,Zimmermann_2016,Liu_2017,Song_2017,Ni_2016,Ni_2018b,Ni_2018a,Kheifets_2020,Eicke_2018,
Douguet_2018,Bray_2018,Crowe_2018,Ren_2018,Tan_2018,Sokolovski_2018,Quan_2019,Douguet_2019,Hofmann_2019,Serov_2019,Wang_2019,Han_2019,Yuan_2019}, confirming or disputing the interpretation of the  experimental attoclock results  \cite{Eckle_2008a,Eckle_2008b,Pfeiffer_2012,Landsman_2014o,Camus_2017,Sainadh_2019}.  The main difficulty  stems from the fact that in such an experiment the photoelectron momentum distribution (PMD) is measured, rather than directly the tunneling time. This time is retrieved using time-to-angle mapping for photoelectrons tunnel-ionized in a laser field of circular polarization (elliptical polarization close to circular)~\cite{Eckle_2008a}. This mapping  straightforwardly follows from the so-called simple man model \cite{Corkum_1993}. According to this the photoelectron emission angle is determined by the direction of the laser vector potential at the moment of the electron appearing in the continuum. However, in a real physical situation the extraction of information on the tunneling time from PMD is not straightforward, because the Coulomb field of the atomic core induces a similar effect in PMD as the tunneling time delay (with respect to the peak of the laser field) and this effect is difficult to account for quantum mechanically with high accuracy. For this reason a semiclassical method was proposed \cite{Eckle_2008a,Eckle_2008b,Pfeiffer_2012}, where the tunneling was treated quantum mechanically, but the further electron motion in the continuum under the simultaneous action of the laser and Coulomb fields, classically.  Moreover, the Coulomb field effect  essentially depends on the tunnel exit coordinate,
which in the quasistatic regime of ionization was calculated including tunnel ionization in parabolic coordinates with induced dipole and Stark shift (TIPIS model) \cite{Pfeiffer_2012}. The semiclassical method was further improved, deriving the initial conditions of the classical propagation via the quantum mechanical Wigner trajectory emerging from the tunneling region \cite{Yakaboylu_2013,Camus_2017}. However, nonadiabaticity of the tunneling ionization renders the quasistatic Wigner theory and related  matched quantum-classical model inaccurate at large Keldysh parameters \cite{Keldysh_1965}.

The numerical solutions of the time-dependent Schr\"odinger equation (TDSE) \cite{Ivanov_2014,Muller_1999,Bauer_2006,Patchkovskii_2016,Torlina_2015,Majety_2017,Sainadh_2019} for the attoclock reproduce the experimental results, yielding confidence that the attoclock PMD features have a single electron origin. However, the numerical results do not contribute much to our understanding of the tunneling time. Not long ago the backpropagation method was proposed to deduce the tunneling time delay from the numerical solution of TDSE \cite{Ni_2016,Ni_2018a,Ni_2018b}. In this method, the asymptotic numerical solution of TDSE is simulated by a classical ensemble, which is backpropagated classically up to the tunnel exit, assuming that quantum features near the tunnel exit  are unimportant.
With the backpropagation method a conclusion was drawn that the average time delay is negligible. However, the obtained negative time delays  of several atomic units (a.u.) were not analyzed in detail,  assuming intuitively that the origin of the negative time cannot be the tunneling. The TDSE numerical results have been also compared to the analytical R-matrix (ARM) theory \cite{Torlina_2015}, which is the state-of-the-art theory of the Coulomb-corrected strong field approximation (SFA) \cite{Popruzhenko_2008a,Popruzhenko_2008b,Torlina_2012,Kaushal_2013}. The comparison revealed that TDSE has a negative time delay with respect to ARM at large laser fields which, however, has been interpreted as a consequence of the bound state depletion and frustrated ionization \cite{Nubbemeyer_2008}.
The problem of the accurate description of subtle features of PMD has been addressed in \cite{Klaiber_2018} within
SFA. A new type of quantum orbit was identified there, corresponding to  under-the-barrier recollisions, where interference with the direct ionization path induces a gentle modification of the asymptotic PMD.

There are basically two opinions in interpreting the attoclock experiment, one claiming negligible (or zero) \cite{Torlina_2015,Eicke_2018,Ni_2018b,Ni_2018a,Sainadh_2019,Kheifets_2020}, and  other -- nonnegligible asymptotic tunneling time delay (ATD) \cite{Pfeiffer_2012,Landsman_2014o,Yakaboylu_2013,Camus_2017,Klaiber_2018}. While ATD is read out from the asymptotic PMD, some also considered theoretically the near exit time delay (ETD) \cite{Yakaboylu_2013,Teeny_2016a,Teeny_2016b,Douguet_2018}, observable in a Gedanken experiment with a so-called virtual detector near the tunnel exit \cite{Feuerstein_2003,Wang_2013}. These two faces of the concept of the tunneling time delay should be clearly distinguished.  ATD  is defined by the time delay of the classical trajectory, which is classically backpropagated from the peak of the photoelectron asymptotic wave function up to the exact point of the tunnel exit. The classical backpropagation  is physically relevant, as the electron dynamics in the continuum is quasiclassical. However, near the tunnel exit the quasiclassical dynamics fails, rendering the notion of the classical trajectory inconsistent. Here the dynamical information can be extracted from the quantum mechanical wave function. In particular, the so-called Wigner trajectory is deduced from the latter, which traces the evolution of the peak of the laser-driven part of the electron wave function during the tunneling ionization and defines the physical ETD.  The Wigner trajectory is in accordance with the classical backpropagation trajectory a few de-Broglie wavelengths away from the tunnel exit. However, near the tunnel exit  the Wigner trajectory deviates strongly from the classical one and shows a positive ETD \cite{Yakaboylu_2013,Han_2019}.
We note that ATD and ETD characterise the  tunneling dynamics from different perspectives: While ATD is related to the attoclock protocol, ETD describes how the classical trajectory emerges from the quantum dynamics of the laser driven atomic electron.

\begin{figure}
  \begin{center}
\includegraphics[width=0.23\textwidth]{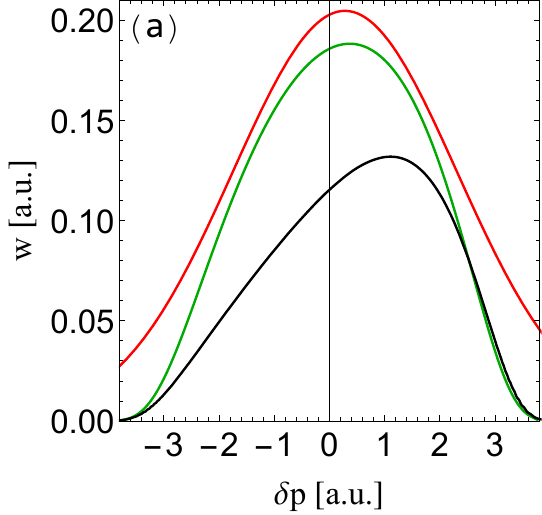}
\includegraphics[width=0.23\textwidth]{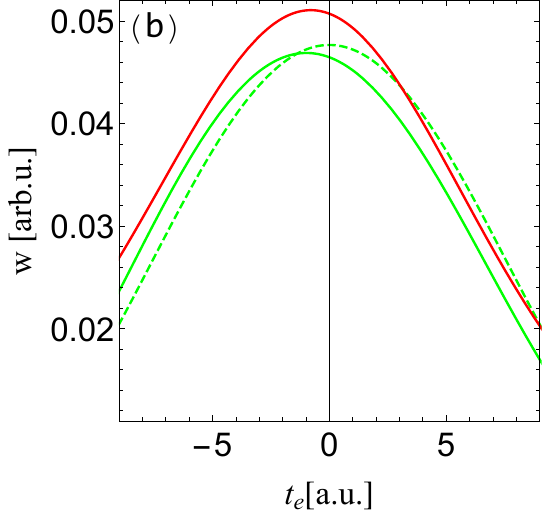}
\includegraphics[width=0.23\textwidth]{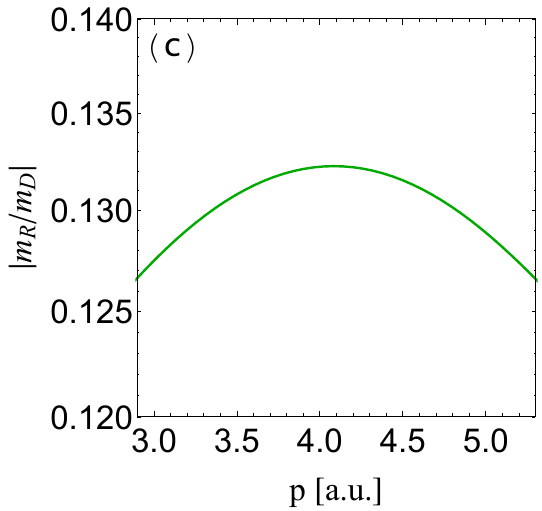}
\includegraphics[width=0.23\textwidth]{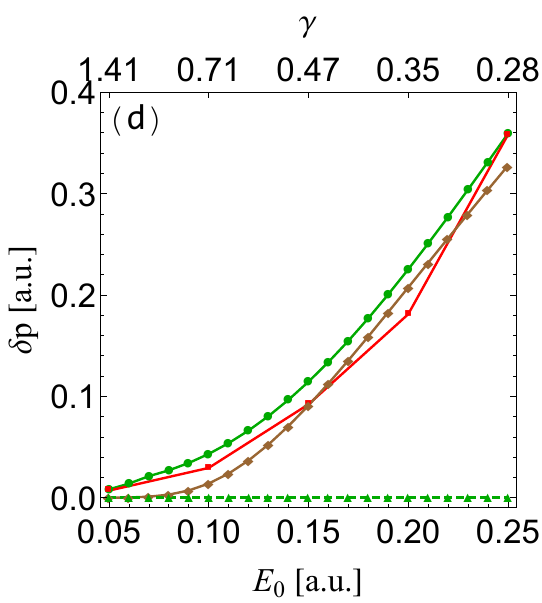}
\caption{ (a) Asymptotic PMD as  function of $\delta p\equiv p+A(0)$ for $E_0=0.25$ a.u.,
where the grid line at $\delta p =0$  shows the PMD peak  with only the direct ionization amplitude $m_D$; (b) Tunneling time distribution  using the backpropagation method of the asymptotic PMD:
(c) The ratio of  the rescattering amplitude to the direct ionization one $|m_R/m_D|$;
(d) The shift of the PMD peak ($\delta p$) due to the tunneling time delay vs the laser field;
    For all panels (green-solid) via the full SFA $m=m_D+m_R$, including sub-barrier direct and rescattered paths, (green-dashed) via  the  SFA direct amplitude, (black) via the numerical TDSE solution, (red)  TDSE with the bound state depletion subtracted \cite{Supplement}, (brown) via the static Wigner trajectory.}
\label{com}
\end{center}
\end{figure}

This Letter is devoted to resolving conflicting conclusions of theories on the tunneling time delay and clarifying the difference of ATD and ETD. We judiciously consider a very basic tunneling scenario which is not plagued with complications stemming from Coulomb effects and the depletion of the bound state, is applicable in the adiabatic as well as in the nonadiabatic regimes, avoids consequent controversial approximations and allows for analytical results. We consider ionization of one-dimensional (1D) atom bound with a zero-range potential driven by an half-cycle laser field. The calculation using SFA \cite{Keldysh_1965,Faisal_1973,Reiss_1980} is carried out fully analytically which facilitates a qualitative comparison  of all differing models. Our results are confirmed by the numerical solutions of TDSE  in 1D as well as in 3D  with linearly and circularly polarized laser pulses. The reasons for the conclusions deviating from Refs.~\cite{Sainadh_2019,Ni_2016,Ni_2018b,Ni_2018a,Kheifets_2020,Eicke_2018} are all analyzed.

We consider ionization of an electron bound in a 1D zero-range potential  $V(x)=-\kappa\delta(x)$, in a half-cycle laser pulse with electric field $ E(t)=-E_0\cos^2(\omega t)$, where $\omega=0.05$ a.u., $\kappa=\sqrt{2I_p}=1$ a.u.
and $I_p$ is the ionization potential. The Keldysh parameter is $\gamma= \tilde{\omega} \kappa/E_0$,  with the effective frequency $\tilde{\omega}\equiv \sqrt{2}\omega$ related to the  $\cos^2$-pulse. Atomic units are used throughout. We employ SFA, with incorporated low-frequency approximation (LFA) for a more accurate treatment  (beyond the Born approximation)
of the recollision \cite{Kroll_1973,Cerkic_2009,Milosevic_2014a}. The LFA validity is justified
as the laser frequency $\omega\ll \varepsilon _r$ \cite{Kroll_1973}, with the recollision energy $\varepsilon _r\sim 1$ a.u. The asymptotic momentum distribution, $w(p)= |m(p)|^2=|m_D(p)+m_R(p)|^2$,  see Fig.~\ref{com}, is determined by interference of  the direct ionization amplitude:
\begin{eqnarray}
m_D(p)=-i\int dt \langle\psi_p^V(t)|H_i(t)|\phi(t)\rangle,
\label{direct}
\end{eqnarray}
and the ionization amplitude with rescattering, described by a second order SFA
\cite{ Becker_2002,Milosevic_2014a}:
\begin{eqnarray}
&&m_R(p)=\label{RRR}\\
&&-\int dt \int_t ds\int dq\langle\psi^V_p(s)|T(p+A(s))|\psi^V_q(s)\rangle\langle \psi^V_q(t)|H_i(t)|\phi(t)\rangle.\nonumber
\end{eqnarray}
Here $\phi(x,t)=\sqrt{\kappa}\exp(-\kappa|x|+i\kappa^2/2t)$ is the bound state wave function,
$\psi^V_p(x,t)$ the Volkov wave function \cite{Volkov_1935}, $H_i(t)=xE(t)$ the electron interaction Hamiltonian with the laser field,  and $A(t)=-\int^t_{t_f}dsE(s)$. In the considered half-cycle laser field the rescattering takes place during the under-the-barrier dynamics, which is in LFA described with the exact laser-free scattering $T$-matrix: $\langle p|T(p)|q \rangle =-(\kappa/2\pi)/(1-i\kappa/|p|)$ \cite{Milosevic_2014a}. The time integration in $m_R$ is  carried out via 3D saddle-point approximation (SPA) \cite{Supplement}. We have also calculated asymptotic PMD via numerical TDSE solution \cite{Supplement}, which is in accordance with the analytical result and shows a momentum shift with respect to the zero-time delay case (PMD via the direct SFA amplitude $m_D$) corresponding to a negative ATD $\delta t=-\delta p/E_0$, see Fig.~\ref{com}.

We retrieve the distribution of ATD from the asymptotic PMD, see  Fig.~\ref{com}(b), using the backpropagating method \cite{Ni_2016}. The classical backpropagation up to the tunnel exit, where the longitudinal velocity is vanishing $p+A(t_e)=0$, is carried out using  the photoelectron asymptotic wave function  $\psi(x,t)=\int m(p) \exp(ipx)dp$, with the total amplitude $m(p)=m_D+m_R$ \cite{Supplement}.
The interference of the direct and the under-the-barrier rescattered trajectories, which is governed by the ratio of amplitudes $m_R/m_D$, induces a visible shift in the asymptotic PMD with respect to the case of the PMD based on the direct trajectory only,  although $|m_R/m_D|\approx 0.13$. For instance, the momentum shift is $\delta p\sim 0.3$~a.u., which is equivalent to the negative ATD $t_e\sim -1$~a.u., at $E_0=0.25$ a.u.,
see Fig.~\ref{com}(b) and (c). The exponential suppression of ATD is governed by the parameter $E_0/E_{th}$ \cite{Klaiber_2018}, with  the threshold field $E_{th}$ of  over-the-barrier ionization (OTBI) \cite{Augst_1989,Supplement}. It is larger near the OTBI threshold  \cite{Yakaboylu_2013,Yakaboylu_2014b} (the shorter the barrier length, the larger is the tunneling time \cite{Peres_1980}).

 In Fig.~\ref{com}(d) the dependence of the momentum shift $\delta p$ due to the tunneling time delay  on the laser field amplitude is shown. Thus, the calculations with our basic tunneling scenario show that the peak of asymptotic PMD can have a time delay up to the order of 1 a.u., see Fig.~\ref{com}(d), due to  interference of the direct and rescattered paths.
 The averaging over PMD decreases the time delay. We can give a simple estimation of the latter property. From \cite{Klaiber_2018}, the negative time delay  is proportional to the Keldysh-exponent $t_e(t)\approx t_{0}\exp\{-2\kappa^3/(3F(t))\}$, with the maximum of the time delay $t_{max}=t_0\exp\{-2\kappa^3/(3E_0)\}$, and $t_0\sim 1/\kappa^2$, such that the averaged time delay can be estimated as
$\langle t_e\rangle\sim \int^{t_f}_{-t_f} t_e(t) w(t)dt/\int^{t_f}_{-t_f} P(t)dt
\sim 0.7 t_{max}$, with the tunneling ionization probability $w(t)\sim \exp\{-2\kappa^3/(3F(t))\}$.

\begin{figure}
\begin{center}	\includegraphics[width=0.4\textwidth]{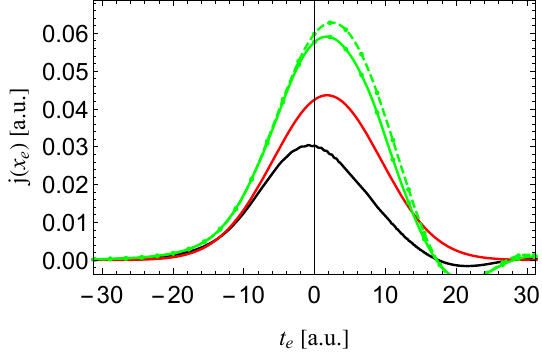}
\caption{ The electron time-dependent current density $j(x_e)$ near the tunnel exit during  tunneling ionization: (green-solid) via the full SFA, and (green-dashed) via the first-order SFA, (black) via the TDSE numerical solution, (red) via the TDSE numerical solution with depletion subtracted \cite{Supplement}. The ETD is positive as exhibited by the SFA curve, as well as by the TDSE with subtracted depletion.  }
      \label{exit}
    \end{center}
 \end{figure}

\begin{figure}[b]
   \begin{center}
   \includegraphics[width=0.35\textwidth]{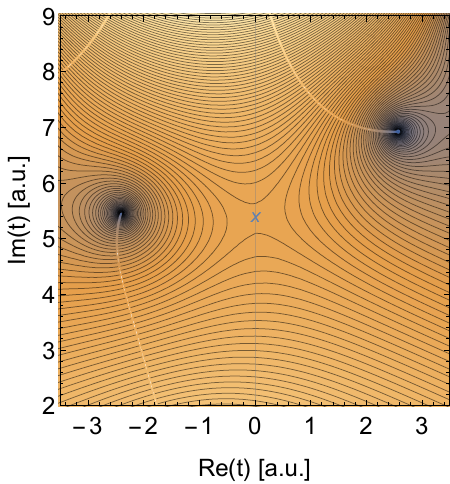}
 \caption{The time-integrand of the ionization amplitude via SFA in the complex plane
     in the case of  the most probable momentum $p=p_{max}$ for the total amplitude $m=m_D+m_R$.
      The saddle-point of the total amplitude (see the cross) is vanishing: ${\rm Re}\{t_s\}=0 $, while ${\rm Re}\{t_s\}<0$ for $m_D$ and $m_R$. The total amplitude has a shifted peak in momentum  due to interference, however, still the same vanishing real parts of the saddle-point at the peak momentum. }
        \label{SP}
  \end{center}
  \end{figure}

The time delay calculated from the second-order SFA in the adiabatic regime $\gamma\ll 1$ is closely related to the static Wigner time delay.  This is illustrated in Fig.~\ref{com}(d), (see brown line), where the estimation of the static Wigner time delay is used, see \cite{Klaiber_2018,Supplement}.
In strong fields, the regime of ionization is adiabatic and the quasistatic Wigner theory is relevant (in Fig.~\ref{com}(d) at $\gamma\lesssim 0.5$). Note that in Ref.~\cite{Camus_2017}, the  deviation of the experimental data from the quasistatic estimation takes place at $\gamma\gtrsim 0.6$.

In Ref.~\cite{Ni_2018b} the average tunneling time delay is calculated for helium ($\kappa=1.345$).
In this case the time delay  is by a factor of $\kappa^2_{He}/\kappa^2\approx 1.8$ smaller than in our $\kappa=1$ a.u. case, since $t_e\sim 1/\kappa^2$~\cite{Klaiber_2013c}. We may compare qualitatively our 1D case of Fig.~\ref{com}(d) ($I_p=0.5)$ with the helium result of Ref.~\cite{Ni_2018b} at the same ratio of $E_0/E^{1D}_{th}=E_0/E^{He}_{th}$, using for helium $ E^{He}_{th}=0.24$ a.u. \cite{Supplement}.
The scaled data of Ref.~\cite{Ni_2018b} provide the average  negative tunneling time delay of the order up to 1 a.u. ($E_0\lesssim 0.25$ a.u.), which qualitatively is in accordance with our model.

With the SFA time-dependent amplitude $m(p,t)$, we retrieve the time-dependent SFA wave function $\Psi (x,t)$ for the ionized electron, and derive ETD from the latter as the peak of the electron current density $j(x_e) $ at the tunnel exit $x_e$:
\begin{eqnarray}
\label{wavefunction}
 \Psi (x,t)&=& \phi(x,t)+\int^\infty_{-\infty} dp\int^t_{-\infty} dt' \psi_p^V(x,t')m(p,t'),
\end{eqnarray}
see also \cite{Klaiber_2013c,Yakaboylu_2013,Canario_2021}. The quantum mechanical description allows to find the physical time delay at the exit  which is read out as the peak of the time-dependent electron current density near the tunnel exit,
see the current density  distribution in Fig.~\ref{exit}, which shows that the most probable ETD is positive. This is also observed in the static tunneling case via the Wigner trajectory $ t^W_e\sim 1/E_0^{2/3}$, yielding also a nonvanishing group velocity at the tunnel exit  $ v_W\sim E_0^{1/3}$ \cite{Klaiber_2013c,Yakaboylu_2013}. The SFA wave function in Fig.~\ref{exit} is calculated via the direct ionization path. The difference between the TDSE and SFA results is due to the contribution of the recolliding trajectory. Thus, Fig.~\ref{exit} shows that ETD is mostly determined by the direct ionization path. The inclusion of the recollision path disturbs  ETD only slightly.
Thus,  the following picture of tunneling ionization emerges: the Wigner trajectory (the peak of the laser-driven part of the wave function) inside the barrier and near the tunnel exit, where bound and free parts of the wave function are inseparable, shows a positive ETD. A few  de-Broglie wavelengths away from the tunnel exit  the most probable classical trajectory emerges from the Wigner one. The classical trajectory shows a negative ATD, which is vanishing in the deep tunneling regime $E_0\ll E_{th}$, however, ETD  is  largest in this regime.

\begin{figure}
\begin{center}
\includegraphics[width=0.5\textwidth]{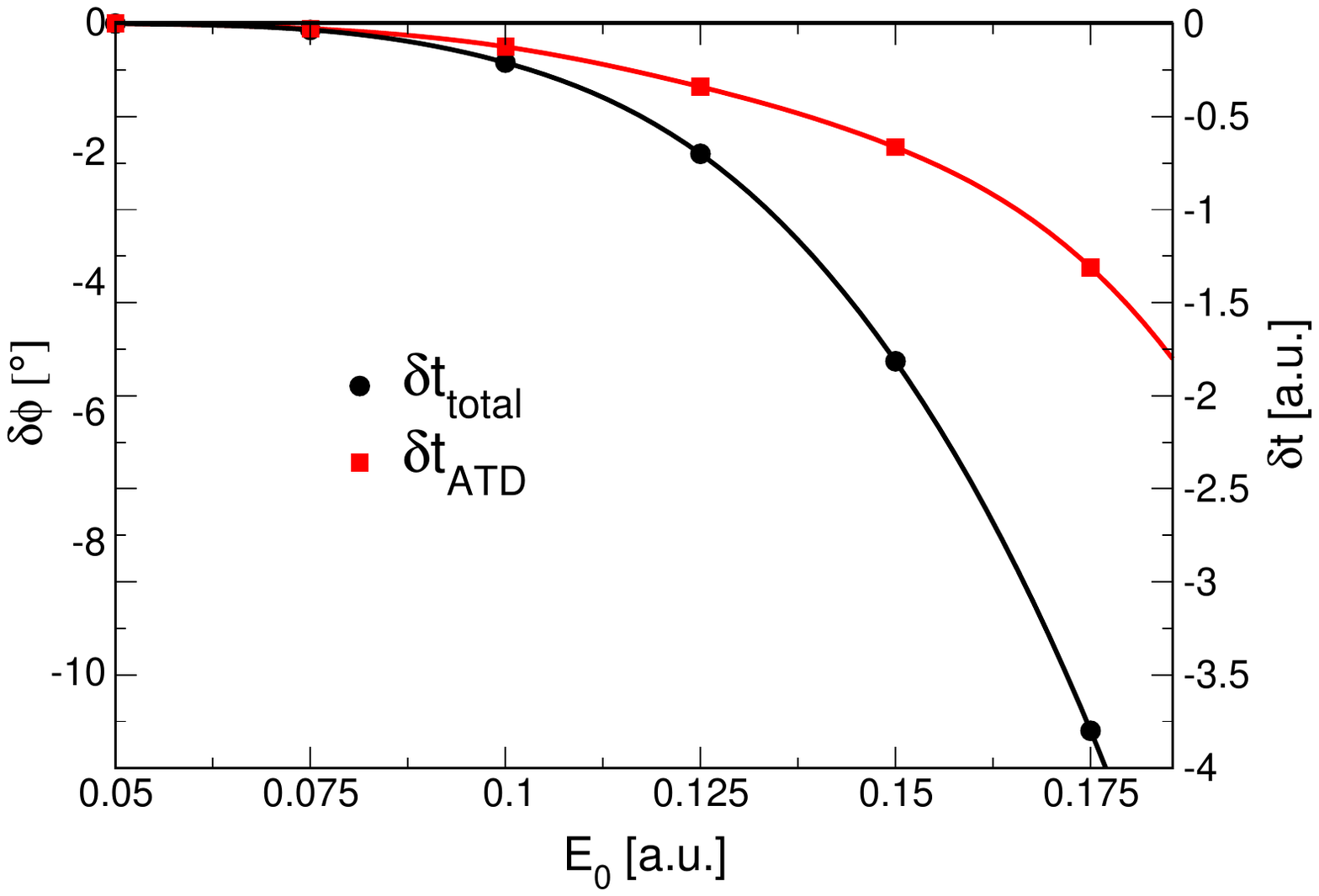}
 \caption{Time delay calculated by the classical backpropagation (BP) of
3D TDSE results with a Yukawa potential
$V(r)=-(Z/r)\exp(-r)$, with  $Z=1.908$ in a circularly polarized
laser pulse: (black circles) total time shift $\delta t_{total}$(BP), and
(red squares) final time shift $\delta t_{ATD}$(BP) with the depletion subtracted. The field strengths, $E_0$, are shown up to $E^{Yukawa}_{th}$.}
       \label{Yukawa}
    \end{center}
 \end{figure}

In Ref.~\cite{Eicke_2018} a  ``trajectory-free" method is proposed to address the tunneling time problem. It is based on the assumption that the (real part of the) saddle-point of the time-integrand of the ionization amplitude determines the ionization time (ATD). We question this assumption. Our line of reasoning is the following. From one side, with our simple SFA model we show that in the near OTBI regime the asymptotic momentum distribution is shifted with respect to the zero ATD case, i.e., demonstrating a nonvanishing ATD. From another side, we apply the ``trajectory-free" method to the SFA wave function \cite{Supplement}, i.e., represent the SFA wave function as a time integral, calculate the time saddle-point, and obtain that the saddle-point of the ionization wave function is zero in the same case, which shows a shift in the momentum distribution  corresponding to the nonzero ATD, see Fig.~\ref{SP}. This demonstrates that the time saddle-point and ATD are not equivalent. We interpret it as follows. The time saddle-point indicates the complex time when the electron ionization path starts at the atomic core. Both amplitudes $m_D$ and $m_R$ start at $Re\{t_s\}=0$ at the core, and each amplitude generates a momentum distribution with the peak corresponding to vanishing ATD. However, their interference causes a deviation of the momentum distribution peak from the no-time-delay model. In short, the time saddle-point is the ionization starting point at the core and it does not coincide with the time-delay because the origin of the latter is the interference of two paths.

Finally, we  have to comment on the numerically calculated  vanishing ATD in  \cite{Sainadh_2019} for an attoclock in the case of a  short-range Yukawa potential $V(r)=-(Z/r)\exp (-r)$, with
$Z=1.908$. The result of vanishing ATD is due to the fact that the range of the field strength of the calculation is not high enough $E_0\lesssim 0.075$. The threshold field of the applied Yukawa potential is $E^{Yukawa}_{th}\approx 0.185$, it 
 is smaller than that of the 1D short-range potential  $E^{Yukawa}_{th}/E^{1D}_{th}\sim 0.7$. From the latter it is expectable to have a sizable ATD, for instance of the order of 1 a.u.,  near the threshold at $E_0= 0.175$, while in the short-range potential it is observed at larger value $E_0= 0.23$, see Fig.~\ref{com}(d). More detailed  $E_0/E_{th}$-scaling is different in 1D and 3D cases, which stems from the wave packet spreading factor in the transverse direction \cite{Klaiber_2018}.
   Furthermore, with the given high charge $Z\approx 2$,  the  momentum transfer at the tunnel exit due to the atomic potential is not negligible at high field strengths, when the tunnel exit is close to the core \cite{Supplement}. The attoclock angular shift due to the atomic potential corresponds to the positive time delay, which counteracts the negative ATD. We have carried out calculations of the ATD for the given Yukawa potential in a large range of field strengths using numerical 3D TDSE solutions for attoclock scenarios \cite{Supplement}. In Fig.~\ref{Yukawa} we show the total time delay calculated via the backpropagation  from the 3D TDSE wave function, as well as the time delay after the subtraction of the depletion contribution. The attoclock offset angle is vanishingly small when the field is not large $E_0\lesssim 0.075$ (up to the laser intensity $4\times 10^{14}$~W/cm$^2$), in agreement with the calculation of \cite{Sainadh_2019}, but increases at high field strengths showing a negative time delay up to $1.75$ a.u.

Concluding, we have analyzed the tunneling time delay in strong field ionization employing a simple tunneling scenario where leading theoretical approaches with seemingly conflicting conclusions could be compared without critical approximations. This way we have demonstrated that the peak of the tunneling wave packet emerging from the barrier around the tunnel exit significantly deviates from the most probable classical backpropagated trajectory, featuring a positive ETD. It however asymptotically merges with the backpropagated trajectory, which itself shows negative ATD, originating from the interference of the direct and recolliding sub-barrier paths. Finally, in explaining the absence of tunneling times from remaining other methods, we have clarified that there are indeed no conflicts among the various approaches.

\bibliography{strong_fields_bibliography}

\begin{thebibliography}{66}%
\makeatletter
\providecommand \@ifxundefined [1]{%
 \@ifx{#1\undefined}
}%
\providecommand \@ifnum [1]{%
 \ifnum #1\expandafter \@firstoftwo
 \else \expandafter \@secondoftwo
 \fi
}%
\providecommand \@ifx [1]{%
 \ifx #1\expandafter \@firstoftwo
 \else \expandafter \@secondoftwo
 \fi
}%
\providecommand \natexlab [1]{#1}%
\providecommand \enquote  [1]{``#1''}%
\providecommand \bibnamefont  [1]{#1}%
\providecommand \bibfnamefont [1]{#1}%
\providecommand \citenamefont [1]{#1}%
\providecommand \href@noop [0]{\@secondoftwo}%
\providecommand \href [0]{\begingroup \@sanitize@url \@href}%
\providecommand \@href[1]{\@@startlink{#1}\@@href}%
\providecommand \@@href[1]{\endgroup#1\@@endlink}%
\providecommand \@sanitize@url [0]{\catcode `\\12\catcode `\$12\catcode
  `\&12\catcode `\#12\catcode `\^12\catcode `\_12\catcode `\%12\relax}%
\providecommand \@@startlink[1]{}%
\providecommand \@@endlink[0]{}%
\providecommand \url  [0]{\begingroup\@sanitize@url \@url }%
\providecommand \@url [1]{\endgroup\@href {#1}{\urlprefix }}%
\providecommand \urlprefix  [0]{URL }%
\providecommand \Eprint [0]{\href }%
\providecommand \doibase [0]{https://doi.org/}%
\providecommand \selectlanguage [0]{\@gobble}%
\providecommand \bibinfo  [0]{\@secondoftwo}%
\providecommand \bibfield  [0]{\@secondoftwo}%
\providecommand \translation [1]{[#1]}%
\providecommand \BibitemOpen [0]{}%
\providecommand \bibitemStop [0]{}%
\providecommand \bibitemNoStop [0]{.\EOS\space}%
\providecommand \EOS [0]{\spacefactor3000\relax}%
\providecommand \BibitemShut  [1]{\csname bibitem#1\endcsname}%
\let\auto@bib@innerbib\@empty
\bibitem [{\citenamefont {Ramos}\ \emph {et~al.}(2020)\citenamefont {Ramos},
  \citenamefont {Spierings}, \citenamefont {Racicot},\ and\ \citenamefont
  {Steinberg}}]{Ramos_2020}%
  \BibitemOpen
  \bibfield  {author} {\bibinfo {author} {\bibfnamefont {R.}~\bibnamefont
  {Ramos}}, \bibinfo {author} {\bibfnamefont {D.}~\bibnamefont {Spierings}},
  \bibinfo {author} {\bibfnamefont {I.}~\bibnamefont {Racicot}},\ and\ \bibinfo
  {author} {\bibfnamefont {A.~M.}\ \bibnamefont {Steinberg}},\ }\bibfield
  {title} {\bibinfo {title} {Measurement of the time spent by a tunnelling atom
  within the barrier region},\ }\href@noop {} {\bibfield  {journal} {\bibinfo
  {journal} {Nature}\ }\textbf {\bibinfo {volume} {583}},\ \bibinfo {pages}
  {529–532} (\bibinfo {year} {2020})}\BibitemShut {NoStop}%
\bibitem [{\citenamefont {Eckle}\ \emph
  {et~al.}(2008{\natexlab{a}})\citenamefont {Eckle}, \citenamefont {Smolarski},
  \citenamefont {Schlup}, \citenamefont {Biegert}, \citenamefont {Staudte},
  \citenamefont {Sch\"offler}, \citenamefont {Muller}, \citenamefont
  {D\"orner},\ and\ \citenamefont {Keller}}]{Eckle_2008a}%
  \BibitemOpen
  \bibfield  {author} {\bibinfo {author} {\bibfnamefont {P.}~\bibnamefont
  {Eckle}}, \bibinfo {author} {\bibfnamefont {M.}~\bibnamefont {Smolarski}},
  \bibinfo {author} {\bibfnamefont {F.}~\bibnamefont {Schlup}}, \bibinfo
  {author} {\bibfnamefont {J.}~\bibnamefont {Biegert}}, \bibinfo {author}
  {\bibfnamefont {A.}~\bibnamefont {Staudte}}, \bibinfo {author} {\bibfnamefont
  {M.}~\bibnamefont {Sch\"offler}}, \bibinfo {author} {\bibfnamefont {H.~G.}\
  \bibnamefont {Muller}}, \bibinfo {author} {\bibfnamefont {R.}~\bibnamefont
  {D\"orner}},\ and\ \bibinfo {author} {\bibfnamefont {U.}~\bibnamefont
  {Keller}},\ }\bibfield  {title} {\bibinfo {title} {Attosecond angular
  streaking},\ }\href@noop {} {\bibfield  {journal} {\bibinfo  {journal}
  {Nature Phys.}\ }\textbf {\bibinfo {volume} {4}},\ \bibinfo {pages} {565}
  (\bibinfo {year} {2008}{\natexlab{a}})}\BibitemShut {NoStop}%
\bibitem [{\citenamefont {Eckle}\ \emph
  {et~al.}(2008{\natexlab{b}})\citenamefont {Eckle}, \citenamefont {Pfeiffer},
  \citenamefont {Cirelli}, \citenamefont {Staudte}, \citenamefont {D\"orner},
  \citenamefont {Muller}, \citenamefont {B\"uttiker},\ and\ \citenamefont
  {Keller}}]{Eckle_2008b}%
  \BibitemOpen
  \bibfield  {author} {\bibinfo {author} {\bibfnamefont {P.}~\bibnamefont
  {Eckle}}, \bibinfo {author} {\bibfnamefont {A.~N.}\ \bibnamefont {Pfeiffer}},
  \bibinfo {author} {\bibfnamefont {C.}~\bibnamefont {Cirelli}}, \bibinfo
  {author} {\bibfnamefont {A.}~\bibnamefont {Staudte}}, \bibinfo {author}
  {\bibfnamefont {R.}~\bibnamefont {D\"orner}}, \bibinfo {author}
  {\bibfnamefont {H.~G.}\ \bibnamefont {Muller}}, \bibinfo {author}
  {\bibfnamefont {M.}~\bibnamefont {B\"uttiker}},\ and\ \bibinfo {author}
  {\bibfnamefont {U.}~\bibnamefont {Keller}},\ }\bibfield  {title} {\bibinfo
  {title} {Attosecond ionization and tunneling delay time measurements in
  helium},\ }\href@noop {} {\bibfield  {journal} {\bibinfo  {journal}
  {Science}\ }\textbf {\bibinfo {volume} {322}},\ \bibinfo {pages} {1525}
  (\bibinfo {year} {2008}{\natexlab{b}})}\BibitemShut {NoStop}%
\bibitem [{\citenamefont {Pfeiffer}\ \emph {et~al.}(2012)\citenamefont
  {Pfeiffer}, \citenamefont {Cirelli}, \citenamefont {Smolarski}, \citenamefont
  {Dimitrovski}, \citenamefont {Abu-samha}, \citenamefont {Madsen},\ and\
  \citenamefont {Keller}}]{Pfeiffer_2012}%
  \BibitemOpen
  \bibfield  {author} {\bibinfo {author} {\bibfnamefont {A.~N.}\ \bibnamefont
  {Pfeiffer}}, \bibinfo {author} {\bibfnamefont {C.}~\bibnamefont {Cirelli}},
  \bibinfo {author} {\bibfnamefont {M.}~\bibnamefont {Smolarski}}, \bibinfo
  {author} {\bibfnamefont {D.}~\bibnamefont {Dimitrovski}}, \bibinfo {author}
  {\bibfnamefont {M.}~\bibnamefont {Abu-samha}}, \bibinfo {author}
  {\bibfnamefont {L.~B.}\ \bibnamefont {Madsen}},\ and\ \bibinfo {author}
  {\bibfnamefont {U.}~\bibnamefont {Keller}},\ }\bibfield  {title} {\bibinfo
  {title} {Attoclock reveals natural coordinates of the laser-induced
  tunnelling current flow in atoms},\ }\href@noop {} {\bibfield  {journal}
  {\bibinfo  {journal} {Nature Phys.}\ }\textbf {\bibinfo {volume} {8}},\
  \bibinfo {pages} {76} (\bibinfo {year} {2012})}\BibitemShut {NoStop}%
\bibitem [{\citenamefont {Landsman}\ \emph {et~al.}(2014)\citenamefont
  {Landsman}, \citenamefont {Weger}, \citenamefont {Maurer}, \citenamefont
  {Boge}, \citenamefont {Ludwig}, \citenamefont {Heuser}, \citenamefont
  {Cirelli}, \citenamefont {Gallmann},\ and\ \citenamefont
  {Keller}}]{Landsman_2014o}%
  \BibitemOpen
  \bibfield  {author} {\bibinfo {author} {\bibfnamefont {A.~S.}\ \bibnamefont
  {Landsman}}, \bibinfo {author} {\bibfnamefont {M.}~\bibnamefont {Weger}},
  \bibinfo {author} {\bibfnamefont {J.}~\bibnamefont {Maurer}}, \bibinfo
  {author} {\bibfnamefont {R.}~\bibnamefont {Boge}}, \bibinfo {author}
  {\bibfnamefont {A.}~\bibnamefont {Ludwig}}, \bibinfo {author} {\bibfnamefont
  {S.}~\bibnamefont {Heuser}}, \bibinfo {author} {\bibfnamefont
  {C.}~\bibnamefont {Cirelli}}, \bibinfo {author} {\bibfnamefont
  {L.}~\bibnamefont {Gallmann}},\ and\ \bibinfo {author} {\bibfnamefont
  {U.}~\bibnamefont {Keller}},\ }\bibfield  {title} {\bibinfo {title}
  {Ultrafast resolution of tunneling delay time},\ }\href@noop {} {\bibfield
  {journal} {\bibinfo  {journal} {Optica}\ }\textbf {\bibinfo {volume} {1}},\
  \bibinfo {pages} {343} (\bibinfo {year} {2014})}\BibitemShut {NoStop}%
\bibitem [{\citenamefont {Camus}\ \emph {et~al.}(2017)\citenamefont {Camus},
  \citenamefont {Yakaboylu}, \citenamefont {Fechner}, \citenamefont {Klaiber},
  \citenamefont {Laux}, \citenamefont {Mi}, \citenamefont {Hatsagortsyan},
  \citenamefont {Pfeifer}, \citenamefont {Keitel},\ and\ \citenamefont
  {Moshammer}}]{Camus_2017}%
  \BibitemOpen
  \bibfield  {author} {\bibinfo {author} {\bibfnamefont {N.}~\bibnamefont
  {Camus}}, \bibinfo {author} {\bibfnamefont {E.}~\bibnamefont {Yakaboylu}},
  \bibinfo {author} {\bibfnamefont {L.}~\bibnamefont {Fechner}}, \bibinfo
  {author} {\bibfnamefont {M.}~\bibnamefont {Klaiber}}, \bibinfo {author}
  {\bibfnamefont {M.}~\bibnamefont {Laux}}, \bibinfo {author} {\bibfnamefont
  {Y.}~\bibnamefont {Mi}}, \bibinfo {author} {\bibfnamefont {K.~Z.}\
  \bibnamefont {Hatsagortsyan}}, \bibinfo {author} {\bibfnamefont
  {T.}~\bibnamefont {Pfeifer}}, \bibinfo {author} {\bibfnamefont {C.~H.}\
  \bibnamefont {Keitel}},\ and\ \bibinfo {author} {\bibfnamefont
  {R.}~\bibnamefont {Moshammer}},\ }\bibfield  {title} {\bibinfo {title}
  {Experimental evidence for wigner's tunneling time},\ }\href@noop {}
  {\bibfield  {journal} {\bibinfo  {journal} {Phys. Rev. Lett.}\ }\textbf
  {\bibinfo {volume} {119}},\ \bibinfo {pages} {023201} (\bibinfo {year}
  {2017})}\BibitemShut {NoStop}%
\bibitem [{\citenamefont {Sainadh}\ \emph {et~al.}(2019)\citenamefont
  {Sainadh}, \citenamefont {Xu}, \citenamefont {Wang}, \citenamefont
  {Atia-Tul-Noor}, \citenamefont {Wallace}, \citenamefont {Douguet},
  \citenamefont {Bray}, \citenamefont {Ivanov}, \citenamefont {Bartschat},
  \citenamefont {Kheifets}, \citenamefont {Sang},\ and\ \citenamefont
  {Litvinyuk}}]{Sainadh_2019}%
  \BibitemOpen
  \bibfield  {author} {\bibinfo {author} {\bibfnamefont {U.~S.}\ \bibnamefont
  {Sainadh}}, \bibinfo {author} {\bibfnamefont {H.}~\bibnamefont {Xu}},
  \bibinfo {author} {\bibfnamefont {X.}~\bibnamefont {Wang}}, \bibinfo {author}
  {\bibfnamefont {A.}~\bibnamefont {Atia-Tul-Noor}}, \bibinfo {author}
  {\bibfnamefont {W.~C.}\ \bibnamefont {Wallace}}, \bibinfo {author}
  {\bibfnamefont {N.}~\bibnamefont {Douguet}}, \bibinfo {author} {\bibfnamefont
  {A.}~\bibnamefont {Bray}}, \bibinfo {author} {\bibfnamefont {I.}~\bibnamefont
  {Ivanov}}, \bibinfo {author} {\bibfnamefont {K.}~\bibnamefont {Bartschat}},
  \bibinfo {author} {\bibfnamefont {A.}~\bibnamefont {Kheifets}}, \bibinfo
  {author} {\bibfnamefont {R.~T.}\ \bibnamefont {Sang}},\ and\ \bibinfo
  {author} {\bibfnamefont {I.~V.}\ \bibnamefont {Litvinyuk}},\ }\bibfield
  {title} {\bibinfo {title} {Attosecond angular streaking and tunnelling time
  in atomic hydrogen},\ }\href@noop {} {\bibfield  {journal} {\bibinfo
  {journal} {Nature}\ }\textbf {\bibinfo {volume} {568}},\ \bibinfo {pages}
  {75} (\bibinfo {year} {2019})}\BibitemShut {NoStop}%
\bibitem [{\citenamefont {Czirj\'ak}\ \emph {et~al.}(2000)\citenamefont
  {Czirj\'ak}, \citenamefont {Kopold}, \citenamefont {Becker}, \citenamefont
  {Kleber},\ and\ \citenamefont {Schleich}}]{Czirjak_2000}%
  \BibitemOpen
  \bibfield  {author} {\bibinfo {author} {\bibfnamefont {A.}~\bibnamefont
  {Czirj\'ak}}, \bibinfo {author} {\bibfnamefont {R.}~\bibnamefont {Kopold}},
  \bibinfo {author} {\bibfnamefont {W.}~\bibnamefont {Becker}}, \bibinfo
  {author} {\bibfnamefont {M.}~\bibnamefont {Kleber}},\ and\ \bibinfo {author}
  {\bibfnamefont {W.}~\bibnamefont {Schleich}},\ }\bibfield  {title} {\bibinfo
  {title} {{The Wigner function for tunneling in a uniform static electric
  field}},\ }\href@noop {} {\bibfield  {journal} {\bibinfo  {journal} {Opt.
  Commun.}\ }\textbf {\bibinfo {volume} {179}},\ \bibinfo {pages} {29 }
  (\bibinfo {year} {2000})}\BibitemShut {NoStop}%
\bibitem [{\citenamefont {Lein}(2011)}]{Lein_2011}%
  \BibitemOpen
  \bibfield  {author} {\bibinfo {author} {\bibfnamefont {M.}~\bibnamefont
  {Lein}},\ }\bibfield  {title} {\bibinfo {title} {Streaking analysis of
  strong-field ionisation},\ }\href@noop {} {\bibfield  {journal} {\bibinfo
  {journal} {J. Mod. Opt.}\ }\textbf {\bibinfo {volume} {58}},\ \bibinfo
  {pages} {1188} (\bibinfo {year} {2011})}\BibitemShut {NoStop}%
\bibitem [{\citenamefont {Orlando}\ \emph
  {et~al.}(2014{\natexlab{a}})\citenamefont {Orlando}, \citenamefont
  {McDonald}, \citenamefont {Protik}, \citenamefont {Vampa},\ and\
  \citenamefont {Brabec}}]{Orlando_2014}%
  \BibitemOpen
  \bibfield  {author} {\bibinfo {author} {\bibfnamefont {G.}~\bibnamefont
  {Orlando}}, \bibinfo {author} {\bibfnamefont {C.~R.}\ \bibnamefont
  {McDonald}}, \bibinfo {author} {\bibfnamefont {N.~H.}\ \bibnamefont
  {Protik}}, \bibinfo {author} {\bibfnamefont {G.}~\bibnamefont {Vampa}},\ and\
  \bibinfo {author} {\bibfnamefont {T.}~\bibnamefont {Brabec}},\ }\bibfield
  {title} {\bibinfo {title} {Tunnelling time, what does it mean?},\ }\href@noop
  {} {\bibfield  {journal} {\bibinfo  {journal} {J. Phys. B}\ }\textbf
  {\bibinfo {volume} {47}},\ \bibinfo {pages} {204002} (\bibinfo {year}
  {2014}{\natexlab{a}})}\BibitemShut {NoStop}%
\bibitem [{\citenamefont {Orlando}\ \emph
  {et~al.}(2014{\natexlab{b}})\citenamefont {Orlando}, \citenamefont
  {McDonald}, \citenamefont {Protik},\ and\ \citenamefont
  {Brabec}}]{Orlando_2014PRA}%
  \BibitemOpen
  \bibfield  {author} {\bibinfo {author} {\bibfnamefont {G.}~\bibnamefont
  {Orlando}}, \bibinfo {author} {\bibfnamefont {C.~R.}\ \bibnamefont
  {McDonald}}, \bibinfo {author} {\bibfnamefont {N.~H.}\ \bibnamefont
  {Protik}},\ and\ \bibinfo {author} {\bibfnamefont {T.}~\bibnamefont
  {Brabec}},\ }\bibfield  {title} {\bibinfo {title} {Identification of the
  keldysh time as a lower limit for the tunneling time},\ }\href@noop {}
  {\bibfield  {journal} {\bibinfo  {journal} {Phys. Rev. A}\ }\textbf {\bibinfo
  {volume} {89}},\ \bibinfo {pages} {014102} (\bibinfo {year}
  {2014}{\natexlab{b}})}\BibitemShut {NoStop}%
\bibitem [{\citenamefont {Landsman}\ and\ \citenamefont
  {Keller}(2014)}]{Landsman_2014}%
  \BibitemOpen
  \bibfield  {author} {\bibinfo {author} {\bibfnamefont {A.~S.}\ \bibnamefont
  {Landsman}}\ and\ \bibinfo {author} {\bibfnamefont {U.}~\bibnamefont
  {Keller}},\ }\bibfield  {title} {\bibinfo {title} {Tunnelling time in strong
  field ionisation},\ }\href@noop {} {\bibfield  {journal} {\bibinfo  {journal}
  {J. Phys. B}\ }\textbf {\bibinfo {volume} {47}},\ \bibinfo {pages} {204024}
  (\bibinfo {year} {2014})}\BibitemShut {NoStop}%
\bibitem [{\citenamefont {Torlina}\ \emph {et~al.}(2015)\citenamefont
  {Torlina}, \citenamefont {Morales}, \citenamefont {Kaushal}, \citenamefont
  {Ivanov}, \citenamefont {Kheifets}, \citenamefont {Zielinski}, \citenamefont
  {Scrinzi}, \citenamefont {Muller}, \citenamefont {Sukiasyan}, \citenamefont
  {Ivanov},\ and\ \citenamefont {Smirnova}}]{Torlina_2015}%
  \BibitemOpen
  \bibfield  {author} {\bibinfo {author} {\bibfnamefont {L.}~\bibnamefont
  {Torlina}}, \bibinfo {author} {\bibfnamefont {F.}~\bibnamefont {Morales}},
  \bibinfo {author} {\bibfnamefont {J.}~\bibnamefont {Kaushal}}, \bibinfo
  {author} {\bibfnamefont {I.}~\bibnamefont {Ivanov}}, \bibinfo {author}
  {\bibfnamefont {A.}~\bibnamefont {Kheifets}}, \bibinfo {author}
  {\bibfnamefont {A.}~\bibnamefont {Zielinski}}, \bibinfo {author}
  {\bibfnamefont {A.}~\bibnamefont {Scrinzi}}, \bibinfo {author} {\bibfnamefont
  {H.~G.}\ \bibnamefont {Muller}}, \bibinfo {author} {\bibfnamefont
  {S.}~\bibnamefont {Sukiasyan}}, \bibinfo {author} {\bibfnamefont
  {M.}~\bibnamefont {Ivanov}},\ and\ \bibinfo {author} {\bibfnamefont
  {O.}~\bibnamefont {Smirnova}},\ }\bibfield  {title} {\bibinfo {title}
  {Interpreting attoclock measurements of tunnelling times},\ }\href@noop {}
  {\bibfield  {journal} {\bibinfo  {journal} {Nat. Phys.}\ }\textbf {\bibinfo
  {volume} {11}},\ \bibinfo {pages} {503} (\bibinfo {year} {2015})}\BibitemShut
  {NoStop}%
\bibitem [{\citenamefont {Landsman}\ and\ \citenamefont
  {Keller}(2015)}]{Landsman_2015}%
  \BibitemOpen
  \bibfield  {author} {\bibinfo {author} {\bibfnamefont {A.~S.}\ \bibnamefont
  {Landsman}}\ and\ \bibinfo {author} {\bibfnamefont {U.}~\bibnamefont
  {Keller}},\ }\bibfield  {title} {\bibinfo {title} {Attosecond science and the
  tunnelling time problem},\ }\href@noop {} {\bibfield  {journal} {\bibinfo
  {journal} {Phys. Rep.}\ }\textbf {\bibinfo {volume} {547}},\ \bibinfo {pages}
  {1} (\bibinfo {year} {2015})}\BibitemShut {NoStop}%
\bibitem [{\citenamefont {Teeny}\ \emph
  {et~al.}(2016{\natexlab{a}})\citenamefont {Teeny}, \citenamefont {Yakaboylu},
  \citenamefont {Bauke},\ and\ \citenamefont {Keitel}}]{Teeny_2016a}%
  \BibitemOpen
  \bibfield  {author} {\bibinfo {author} {\bibfnamefont {N.}~\bibnamefont
  {Teeny}}, \bibinfo {author} {\bibfnamefont {E.}~\bibnamefont {Yakaboylu}},
  \bibinfo {author} {\bibfnamefont {H.}~\bibnamefont {Bauke}},\ and\ \bibinfo
  {author} {\bibfnamefont {C.~H.}\ \bibnamefont {Keitel}},\ }\bibfield  {title}
  {\bibinfo {title} {{Ionization Time and Exit Momentum in Strong-Field Tunnel
  Ionization}},\ }\href@noop {} {\bibfield  {journal} {\bibinfo  {journal}
  {Phys. Rev. Lett.}\ }\textbf {\bibinfo {volume} {116}},\ \bibinfo {pages}
  {063003} (\bibinfo {year} {2016}{\natexlab{a}})}\BibitemShut {NoStop}%
\bibitem [{\citenamefont {Teeny}\ \emph
  {et~al.}(2016{\natexlab{b}})\citenamefont {Teeny}, \citenamefont {Keitel},\
  and\ \citenamefont {Bauke}}]{Teeny_2016b}%
  \BibitemOpen
  \bibfield  {author} {\bibinfo {author} {\bibfnamefont {N.}~\bibnamefont
  {Teeny}}, \bibinfo {author} {\bibfnamefont {C.~H.}\ \bibnamefont {Keitel}},\
  and\ \bibinfo {author} {\bibfnamefont {H.}~\bibnamefont {Bauke}},\ }\bibfield
   {title} {\bibinfo {title} {{Virtual-detector approach to tunnel ionization
  and tunneling times}},\ }\href@noop {} {\bibfield  {journal} {\bibinfo
  {journal} {Phys. Rev. A}\ }\textbf {\bibinfo {volume} {94}},\ \bibinfo
  {pages} {022104} (\bibinfo {year} {2016}{\natexlab{b}})}\BibitemShut
  {NoStop}%
\bibitem [{\citenamefont {Yakaboylu}\ \emph {et~al.}(2013)\citenamefont
  {Yakaboylu}, \citenamefont {Klaiber}, \citenamefont {Bauke}, \citenamefont
  {Hatsagortsyan},\ and\ \citenamefont {Keitel}}]{Yakaboylu_2013}%
  \BibitemOpen
  \bibfield  {author} {\bibinfo {author} {\bibfnamefont {E.}~\bibnamefont
  {Yakaboylu}}, \bibinfo {author} {\bibfnamefont {M.}~\bibnamefont {Klaiber}},
  \bibinfo {author} {\bibfnamefont {H.}~\bibnamefont {Bauke}}, \bibinfo
  {author} {\bibfnamefont {K.~Z.}\ \bibnamefont {Hatsagortsyan}},\ and\
  \bibinfo {author} {\bibfnamefont {C.~H.}\ \bibnamefont {Keitel}},\ }\bibfield
   {title} {\bibinfo {title} {Relativistic features and time delay of
  laser-induced tunnel ionization},\ }\href@noop {} {\bibfield  {journal}
  {\bibinfo  {journal} {Phys. Rev. A}\ }\textbf {\bibinfo {volume} {88}},\
  \bibinfo {pages} {063421} (\bibinfo {year} {2013})}\BibitemShut {NoStop}%
\bibitem [{\citenamefont {Klaiber}\ \emph {et~al.}(2018)\citenamefont
  {Klaiber}, \citenamefont {Hatsagortsyan},\ and\ \citenamefont
  {Keitel}}]{Klaiber_2018}%
  \BibitemOpen
  \bibfield  {author} {\bibinfo {author} {\bibfnamefont {M.}~\bibnamefont
  {Klaiber}}, \bibinfo {author} {\bibfnamefont {K.~Z.}\ \bibnamefont
  {Hatsagortsyan}},\ and\ \bibinfo {author} {\bibfnamefont {C.~H.}\
  \bibnamefont {Keitel}},\ }\bibfield  {title} {\bibinfo {title}
  {Under-the-tunneling-barrier recollisions in strong-field ionization},\
  }\href@noop {} {\bibfield  {journal} {\bibinfo  {journal} {Phys. Rev. Lett.}\
  }\textbf {\bibinfo {volume} {120}},\ \bibinfo {pages} {013201} (\bibinfo
  {year} {2018})}\BibitemShut {NoStop}%
\bibitem [{\citenamefont {Zimmermann}\ \emph {et~al.}(2016)\citenamefont
  {Zimmermann}, \citenamefont {Mishra}, \citenamefont {Doran}, \citenamefont
  {Gordon},\ and\ \citenamefont {Landsman}}]{Zimmermann_2016}%
  \BibitemOpen
  \bibfield  {author} {\bibinfo {author} {\bibfnamefont {T.}~\bibnamefont
  {Zimmermann}}, \bibinfo {author} {\bibfnamefont {S.}~\bibnamefont {Mishra}},
  \bibinfo {author} {\bibfnamefont {B.~R.}\ \bibnamefont {Doran}}, \bibinfo
  {author} {\bibfnamefont {D.~F.}\ \bibnamefont {Gordon}},\ and\ \bibinfo
  {author} {\bibfnamefont {A.~S.}\ \bibnamefont {Landsman}},\ }\bibfield
  {title} {\bibinfo {title} {Tunneling time and weak measurement in strong
  field ionization},\ }\href@noop {} {\bibfield  {journal} {\bibinfo  {journal}
  {Phys. Rev. Lett.}\ }\textbf {\bibinfo {volume} {116}},\ \bibinfo {pages}
  {233603} (\bibinfo {year} {2016})}\BibitemShut {NoStop}%
\bibitem [{\citenamefont {Liu}\ \emph {et~al.}(2017)\citenamefont {Liu},
  \citenamefont {Fu}, \citenamefont {Chen}, \citenamefont {Lü}, \citenamefont
  {Zhao}, \citenamefont {Yuan},\ and\ \citenamefont {Zhao}}]{Liu_2017}%
  \BibitemOpen
  \bibfield  {author} {\bibinfo {author} {\bibfnamefont {J.}~\bibnamefont
  {Liu}}, \bibinfo {author} {\bibfnamefont {Y.}~\bibnamefont {Fu}}, \bibinfo
  {author} {\bibfnamefont {W.}~\bibnamefont {Chen}}, \bibinfo {author}
  {\bibfnamefont {Z.}~\bibnamefont {Lü}}, \bibinfo {author} {\bibfnamefont
  {J.}~\bibnamefont {Zhao}}, \bibinfo {author} {\bibfnamefont {J.}~\bibnamefont
  {Yuan}},\ and\ \bibinfo {author} {\bibfnamefont {Z.}~\bibnamefont {Zhao}},\
  }\bibfield  {title} {\bibinfo {title} {Offset angles of photocurrents
  generated in few-cycle circularly polarized laser fields},\ }\href
  {https://doi.org/10.1088/1361-6455/aa575b} {\bibfield  {journal} {\bibinfo
  {journal} {J. Phys. B}\ }\textbf {\bibinfo {volume} {50}},\ \bibinfo {pages}
  {055602} (\bibinfo {year} {2017})}\BibitemShut {NoStop}%
\bibitem [{\citenamefont {Song}\ \emph {et~al.}(2017)\citenamefont {Song},
  \citenamefont {Yang}, \citenamefont {Guo},\ and\ \citenamefont
  {Li}}]{Song_2017}%
  \BibitemOpen
  \bibfield  {author} {\bibinfo {author} {\bibfnamefont {Y.}~\bibnamefont
  {Song}}, \bibinfo {author} {\bibfnamefont {Y.}~\bibnamefont {Yang}}, \bibinfo
  {author} {\bibfnamefont {F.}~\bibnamefont {Guo}},\ and\ \bibinfo {author}
  {\bibfnamefont {S.}~\bibnamefont {Li}},\ }\bibfield  {title} {\bibinfo
  {title} {Revisiting the time-dependent ionization process through the
  bohmian-mechanics method},\ }\href@noop {} {\bibfield  {journal} {\bibinfo
  {journal} {J. Phys. B}\ }\textbf {\bibinfo {volume} {50}},\ \bibinfo {pages}
  {095003} (\bibinfo {year} {2017})}\BibitemShut {NoStop}%
\bibitem [{\citenamefont {Ni}\ \emph {et~al.}(2016)\citenamefont {Ni},
  \citenamefont {Saalmann},\ and\ \citenamefont {Rost}}]{Ni_2016}%
  \BibitemOpen
  \bibfield  {author} {\bibinfo {author} {\bibfnamefont {H.}~\bibnamefont
  {Ni}}, \bibinfo {author} {\bibfnamefont {U.}~\bibnamefont {Saalmann}},\ and\
  \bibinfo {author} {\bibfnamefont {J.-M.}\ \bibnamefont {Rost}},\ }\bibfield
  {title} {\bibinfo {title} {Tunneling ionization time resolved by
  backpropagation},\ }\href@noop {} {\bibfield  {journal} {\bibinfo  {journal}
  {Phys. Rev. Lett.}\ }\textbf {\bibinfo {volume} {117}},\ \bibinfo {pages}
  {023002} (\bibinfo {year} {2016})}\BibitemShut {NoStop}%
\bibitem [{\citenamefont {Ni}\ \emph {et~al.}(2018{\natexlab{a}})\citenamefont
  {Ni}, \citenamefont {Saalmann},\ and\ \citenamefont {Rost}}]{Ni_2018b}%
  \BibitemOpen
  \bibfield  {author} {\bibinfo {author} {\bibfnamefont {H.}~\bibnamefont
  {Ni}}, \bibinfo {author} {\bibfnamefont {U.}~\bibnamefont {Saalmann}},\ and\
  \bibinfo {author} {\bibfnamefont {J.-M.}\ \bibnamefont {Rost}},\ }\bibfield
  {title} {\bibinfo {title} {{Tunneling exit characteristics from classical
  backpropagation of an ionized electron wave packet}},\ }\href@noop {}
  {\bibfield  {journal} {\bibinfo  {journal} {Phys. Rev. A}\ }\textbf {\bibinfo
  {volume} {97}},\ \bibinfo {pages} {013426} (\bibinfo {year}
  {2018}{\natexlab{a}})}\BibitemShut {NoStop}%
\bibitem [{\citenamefont {Ni}\ \emph {et~al.}(2018{\natexlab{b}})\citenamefont
  {Ni}, \citenamefont {Eicke}, \citenamefont {Ruiz}, \citenamefont {Cai},
  \citenamefont {Oppermann}, \citenamefont {Shvetsov-Shilovski},\ and\
  \citenamefont {Pi}}]{Ni_2018a}%
  \BibitemOpen
  \bibfield  {author} {\bibinfo {author} {\bibfnamefont {H.}~\bibnamefont
  {Ni}}, \bibinfo {author} {\bibfnamefont {N.}~\bibnamefont {Eicke}}, \bibinfo
  {author} {\bibfnamefont {C.}~\bibnamefont {Ruiz}}, \bibinfo {author}
  {\bibfnamefont {J.}~\bibnamefont {Cai}}, \bibinfo {author} {\bibfnamefont
  {F.}~\bibnamefont {Oppermann}}, \bibinfo {author} {\bibfnamefont {N.~I.}\
  \bibnamefont {Shvetsov-Shilovski}},\ and\ \bibinfo {author} {\bibfnamefont
  {L.-W.}\ \bibnamefont {Pi}},\ }\bibfield  {title} {\bibinfo {title}
  {{Tunneling criteria and a nonadiabatic term for strong-field ionization}},\
  }\href@noop {} {\bibfield  {journal} {\bibinfo  {journal} {Phys. Rev. A}\
  }\textbf {\bibinfo {volume} {98}},\ \bibinfo {pages} {013411} (\bibinfo
  {year} {2018}{\natexlab{b}})}\BibitemShut {NoStop}%
\bibitem [{\citenamefont {Kheifets}(2020)}]{Kheifets_2020}%
  \BibitemOpen
  \bibfield  {author} {\bibinfo {author} {\bibfnamefont {A.~S.}\ \bibnamefont
  {Kheifets}},\ }\bibfield  {title} {\bibinfo {title} {The attoclock and the
  tunneling time debate},\ }\href@noop {} {\bibfield  {journal} {\bibinfo
  {journal} {J. Phys. B}\ }\textbf {\bibinfo {volume} {53}},\ \bibinfo {pages}
  {072001} (\bibinfo {year} {2020})}\BibitemShut {NoStop}%
\bibitem [{\citenamefont {Eicke}\ and\ \citenamefont
  {Lein}(2018)}]{Eicke_2018}%
  \BibitemOpen
  \bibfield  {author} {\bibinfo {author} {\bibfnamefont {N.}~\bibnamefont
  {Eicke}}\ and\ \bibinfo {author} {\bibfnamefont {M.}~\bibnamefont {Lein}},\
  }\bibfield  {title} {\bibinfo {title} {{Trajectory-free ionization times in
  strong-field ionization}},\ }\href@noop {} {\bibfield  {journal} {\bibinfo
  {journal} {Phys. Rev. A}\ }\textbf {\bibinfo {volume} {97}},\ \bibinfo
  {pages} {031402} (\bibinfo {year} {2018})}\BibitemShut {NoStop}%
\bibitem [{\citenamefont {Douguet}\ and\ \citenamefont
  {Bartschat}(2018)}]{Douguet_2018}%
  \BibitemOpen
  \bibfield  {author} {\bibinfo {author} {\bibfnamefont {N.}~\bibnamefont
  {Douguet}}\ and\ \bibinfo {author} {\bibfnamefont {K.}~\bibnamefont
  {Bartschat}},\ }\bibfield  {title} {\bibinfo {title} {{Dynamics of Tunneling
  Ionization using Bohmian Mechanics}},\ }\href@noop {} {\bibfield  {journal}
  {\bibinfo  {journal} {Phys. Rev. A}\ }\textbf {\bibinfo {volume} {97}},\
  \bibinfo {pages} {013402} (\bibinfo {year} {2018})}\BibitemShut {NoStop}%
\bibitem [{\citenamefont {Bray}\ \emph {et~al.}(2018)\citenamefont {Bray},
  \citenamefont {Eckart},\ and\ \citenamefont {Kheifets}}]{Bray_2018}%
  \BibitemOpen
  \bibfield  {author} {\bibinfo {author} {\bibfnamefont {A.~W.}\ \bibnamefont
  {Bray}}, \bibinfo {author} {\bibfnamefont {S.}~\bibnamefont {Eckart}},\ and\
  \bibinfo {author} {\bibfnamefont {A.~S.}\ \bibnamefont {Kheifets}},\
  }\bibfield  {title} {\bibinfo {title} {Keldysh-rutherford model for the
  attoclock},\ }\href@noop {} {\bibfield  {journal} {\bibinfo  {journal} {Phys.
  Rev. Lett.}\ }\textbf {\bibinfo {volume} {121}},\ \bibinfo {pages} {123201}
  (\bibinfo {year} {2018})}\BibitemShut {NoStop}%
\bibitem [{\citenamefont {Bayta\c{s}}\ \emph {et~al.}(2018)\citenamefont
  {Bayta\c{s}}, \citenamefont {Bojowald},\ and\ \citenamefont
  {Crowe}}]{Crowe_2018}%
  \BibitemOpen
  \bibfield  {author} {\bibinfo {author} {\bibfnamefont {B.}~\bibnamefont
  {Bayta\c{s}}}, \bibinfo {author} {\bibfnamefont {M.}~\bibnamefont
  {Bojowald}},\ and\ \bibinfo {author} {\bibfnamefont {S.}~\bibnamefont
  {Crowe}},\ }\bibfield  {title} {\bibinfo {title} {Canonical tunneling time in
  ionization experiments},\ }\href@noop {} {\bibfield  {journal} {\bibinfo
  {journal} {Phys. Rev. A}\ }\textbf {\bibinfo {volume} {98}},\ \bibinfo
  {pages} {063417} (\bibinfo {year} {2018})}\BibitemShut {NoStop}%
\bibitem [{\citenamefont {Ren}\ \emph {et~al.}(2018)\citenamefont {Ren},
  \citenamefont {Wu}, \citenamefont {Wang},\ and\ \citenamefont
  {Zheng}}]{Ren_2018}%
  \BibitemOpen
  \bibfield  {author} {\bibinfo {author} {\bibfnamefont {X.}~\bibnamefont
  {Ren}}, \bibinfo {author} {\bibfnamefont {Y.}~\bibnamefont {Wu}}, \bibinfo
  {author} {\bibfnamefont {L.}~\bibnamefont {Wang}},\ and\ \bibinfo {author}
  {\bibfnamefont {Y.}~\bibnamefont {Zheng}},\ }\bibfield  {title} {\bibinfo
  {title} {Entangled trajectories during ionization of an h atom driven by
  n-cycle laser pulse},\ }\href@noop {} {\bibfield  {journal} {\bibinfo
  {journal} {Phys. Lett. A}\ }\textbf {\bibinfo {volume} {382}},\ \bibinfo
  {pages} {2662 } (\bibinfo {year} {2018})}\BibitemShut {NoStop}%
\bibitem [{\citenamefont {Tan}\ \emph {et~al.}(2018)\citenamefont {Tan},
  \citenamefont {Zhou}, \citenamefont {He}, \citenamefont {Chen}, \citenamefont
  {Ke}, \citenamefont {Liang}, \citenamefont {Zhu}, \citenamefont {Li},\ and\
  \citenamefont {Lu}}]{Tan_2018}%
  \BibitemOpen
  \bibfield  {author} {\bibinfo {author} {\bibfnamefont {J.}~\bibnamefont
  {Tan}}, \bibinfo {author} {\bibfnamefont {Y.}~\bibnamefont {Zhou}}, \bibinfo
  {author} {\bibfnamefont {M.}~\bibnamefont {He}}, \bibinfo {author}
  {\bibfnamefont {Y.}~\bibnamefont {Chen}}, \bibinfo {author} {\bibfnamefont
  {Q.}~\bibnamefont {Ke}}, \bibinfo {author} {\bibfnamefont {J.}~\bibnamefont
  {Liang}}, \bibinfo {author} {\bibfnamefont {X.}~\bibnamefont {Zhu}}, \bibinfo
  {author} {\bibfnamefont {M.}~\bibnamefont {Li}},\ and\ \bibinfo {author}
  {\bibfnamefont {P.}~\bibnamefont {Lu}},\ }\bibfield  {title} {\bibinfo
  {title} {Determination of the ionization time using attosecond photoelectron
  interferometry},\ }\href@noop {} {\bibfield  {journal} {\bibinfo  {journal}
  {Phys. Rev. Lett.}\ }\textbf {\bibinfo {volume} {121}},\ \bibinfo {pages}
  {253203} (\bibinfo {year} {2018})}\BibitemShut {NoStop}%
\bibitem [{\citenamefont {Sokolovski}\ and\ \citenamefont
  {Akhmatskaya}(2018)}]{Sokolovski_2018}%
  \BibitemOpen
  \bibfield  {author} {\bibinfo {author} {\bibfnamefont {D.}~\bibnamefont
  {Sokolovski}}\ and\ \bibinfo {author} {\bibfnamefont {E.}~\bibnamefont
  {Akhmatskaya}},\ }\bibfield  {title} {\bibinfo {title} {{No time at the end
  of the tunnel}},\ }\href@noop {} {\bibfield  {journal} {\bibinfo  {journal}
  {Commun. Phys.}\ }\textbf {\bibinfo {volume} {1}},\ \bibinfo {pages} {47}
  (\bibinfo {year} {2018})}\BibitemShut {NoStop}%
\bibitem [{\citenamefont {Quan}\ \emph {et~al.}(2019)\citenamefont {Quan},
  \citenamefont {Serov}, \citenamefont {Wei}, \citenamefont {Zhao},
  \citenamefont {Zhou}, \citenamefont {Wang}, \citenamefont {Lai},
  \citenamefont {Kheifets},\ and\ \citenamefont {Liu}}]{Quan_2019}%
  \BibitemOpen
  \bibfield  {author} {\bibinfo {author} {\bibfnamefont {W.}~\bibnamefont
  {Quan}}, \bibinfo {author} {\bibfnamefont {V.~V.}\ \bibnamefont {Serov}},
  \bibinfo {author} {\bibfnamefont {M.}~\bibnamefont {Wei}}, \bibinfo {author}
  {\bibfnamefont {M.}~\bibnamefont {Zhao}}, \bibinfo {author} {\bibfnamefont
  {Y.}~\bibnamefont {Zhou}}, \bibinfo {author} {\bibfnamefont {Y.}~\bibnamefont
  {Wang}}, \bibinfo {author} {\bibfnamefont {X.}~\bibnamefont {Lai}}, \bibinfo
  {author} {\bibfnamefont {A.~S.}\ \bibnamefont {Kheifets}},\ and\ \bibinfo
  {author} {\bibfnamefont {X.}~\bibnamefont {Liu}},\ }\bibfield  {title}
  {\bibinfo {title} {Attosecond molecular angular streaking with all-ionic
  fragments detection},\ }\href@noop {} {\bibfield  {journal} {\bibinfo
  {journal} {Phys. Rev. Lett.}\ }\textbf {\bibinfo {volume} {123}},\ \bibinfo
  {pages} {223204} (\bibinfo {year} {2019})}\BibitemShut {NoStop}%
\bibitem [{\citenamefont {Douguet}\ and\ \citenamefont
  {Bartschat}(2019)}]{Douguet_2019}%
  \BibitemOpen
  \bibfield  {author} {\bibinfo {author} {\bibfnamefont {N.}~\bibnamefont
  {Douguet}}\ and\ \bibinfo {author} {\bibfnamefont {K.}~\bibnamefont
  {Bartschat}},\ }\bibfield  {title} {\bibinfo {title} {Attoclock setup with
  negative ions: A possibility for experimental validation},\ }\href@noop {}
  {\bibfield  {journal} {\bibinfo  {journal} {Phys. Rev. A}\ }\textbf {\bibinfo
  {volume} {99}},\ \bibinfo {pages} {023417} (\bibinfo {year}
  {2019})}\BibitemShut {NoStop}%
\bibitem [{\citenamefont {Hofmann}\ \emph {et~al.}(2019)\citenamefont
  {Hofmann}, \citenamefont {Landsman},\ and\ \citenamefont
  {Keller}}]{Hofmann_2019}%
  \BibitemOpen
  \bibfield  {author} {\bibinfo {author} {\bibfnamefont {C.}~\bibnamefont
  {Hofmann}}, \bibinfo {author} {\bibfnamefont {A.~S.}\ \bibnamefont
  {Landsman}},\ and\ \bibinfo {author} {\bibfnamefont {U.}~\bibnamefont
  {Keller}},\ }\bibfield  {title} {\bibinfo {title} {{Attoclock revisited on
  electron tunnelling time}},\ }\href@noop {} {\bibfield  {journal} {\bibinfo
  {journal} {J. Mod. Opt.}\ }\textbf {\bibinfo {volume} {66}},\ \bibinfo
  {pages} {1052} (\bibinfo {year} {2019})}\BibitemShut {NoStop}%
\bibitem [{\citenamefont {Serov}\ \emph {et~al.}(2019)\citenamefont {Serov},
  \citenamefont {Bray},\ and\ \citenamefont {Kheifets}}]{Serov_2019}%
  \BibitemOpen
  \bibfield  {author} {\bibinfo {author} {\bibfnamefont {V.~V.}\ \bibnamefont
  {Serov}}, \bibinfo {author} {\bibfnamefont {A.~W.}\ \bibnamefont {Bray}},\
  and\ \bibinfo {author} {\bibfnamefont {A.~S.}\ \bibnamefont {Kheifets}},\
  }\bibfield  {title} {\bibinfo {title} {Numerical attoclock on atomic and
  molecular hydrogen},\ }\href@noop {} {\bibfield  {journal} {\bibinfo
  {journal} {Phys. Rev. A}\ }\textbf {\bibinfo {volume} {99}},\ \bibinfo
  {pages} {063428} (\bibinfo {year} {2019})}\BibitemShut {NoStop}%
\bibitem [{\citenamefont {Wang}\ \emph {et~al.}(2019)\citenamefont {Wang},
  \citenamefont {Zhang}, \citenamefont {Li}, \citenamefont {Xu}, \citenamefont
  {Cao}, \citenamefont {Zhou}, \citenamefont {Cao},\ and\ \citenamefont
  {Lu}}]{Wang_2019}%
  \BibitemOpen
  \bibfield  {author} {\bibinfo {author} {\bibfnamefont {R.}~\bibnamefont
  {Wang}}, \bibinfo {author} {\bibfnamefont {Q.}~\bibnamefont {Zhang}},
  \bibinfo {author} {\bibfnamefont {D.}~\bibnamefont {Li}}, \bibinfo {author}
  {\bibfnamefont {S.}~\bibnamefont {Xu}}, \bibinfo {author} {\bibfnamefont
  {P.}~\bibnamefont {Cao}}, \bibinfo {author} {\bibfnamefont {Y.}~\bibnamefont
  {Zhou}}, \bibinfo {author} {\bibfnamefont {W.}~\bibnamefont {Cao}},\ and\
  \bibinfo {author} {\bibfnamefont {P.}~\bibnamefont {Lu}},\ }\bibfield
  {title} {\bibinfo {title} {Identification of tunneling and multiphoton
  ionization in intermediate keldysh parameter regime},\ }\href
  {http://www.opticsexpress.org/abstract.cfm?URI=oe-27-5-6471} {\bibfield
  {journal} {\bibinfo  {journal} {Opt. Express}\ }\textbf {\bibinfo {volume}
  {27}},\ \bibinfo {pages} {6471} (\bibinfo {year} {2019})}\BibitemShut
  {NoStop}%
\bibitem [{\citenamefont {Han}\ \emph {et~al.}(2019)\citenamefont {Han},
  \citenamefont {Ge}, \citenamefont {Fang}, \citenamefont {Yu}, \citenamefont
  {Guo}, \citenamefont {Ma}, \citenamefont {Deng}, \citenamefont {Gong},\ and\
  \citenamefont {Liu}}]{Han_2019}%
  \BibitemOpen
  \bibfield  {author} {\bibinfo {author} {\bibfnamefont {M.}~\bibnamefont
  {Han}}, \bibinfo {author} {\bibfnamefont {P.}~\bibnamefont {Ge}}, \bibinfo
  {author} {\bibfnamefont {Y.}~\bibnamefont {Fang}}, \bibinfo {author}
  {\bibfnamefont {X.}~\bibnamefont {Yu}}, \bibinfo {author} {\bibfnamefont
  {Z.}~\bibnamefont {Guo}}, \bibinfo {author} {\bibfnamefont {X.}~\bibnamefont
  {Ma}}, \bibinfo {author} {\bibfnamefont {Y.}~\bibnamefont {Deng}}, \bibinfo
  {author} {\bibfnamefont {Q.}~\bibnamefont {Gong}},\ and\ \bibinfo {author}
  {\bibfnamefont {Y.}~\bibnamefont {Liu}},\ }\bibfield  {title} {\bibinfo
  {title} {Unifying tunneling pictures of strong-field ionization with an
  improved attoclock},\ }\href@noop {} {\bibfield  {journal} {\bibinfo
  {journal} {Phys. Rev. Lett.}\ }\textbf {\bibinfo {volume} {123}},\ \bibinfo
  {pages} {073201} (\bibinfo {year} {2019})}\BibitemShut {NoStop}%
\bibitem [{\citenamefont {Yuan}(2019)}]{Yuan_2019}%
  \BibitemOpen
  \bibfield  {author} {\bibinfo {author} {\bibfnamefont {M.}~\bibnamefont
  {Yuan}},\ }\bibfield  {title} {\bibinfo {title} {Direct probing of tunneling
  time in strong-field ionization processes by time-dependent wave packets},\
  }\href@noop {} {\bibfield  {journal} {\bibinfo  {journal} {Opt. Express}\
  }\textbf {\bibinfo {volume} {27}},\ \bibinfo {pages} {6502} (\bibinfo {year}
  {2019})}\BibitemShut {NoStop}%
\bibitem [{\citenamefont {Corkum}(1993)}]{Corkum_1993}%
  \BibitemOpen
  \bibfield  {author} {\bibinfo {author} {\bibfnamefont {P.~B.}\ \bibnamefont
  {Corkum}},\ }\href@noop {} {\bibfield  {journal} {\bibinfo  {journal} {Phys.
  Rev. Lett.}\ }\textbf {\bibinfo {volume} {71}},\ \bibinfo {pages} {1994}
  (\bibinfo {year} {1993})}\BibitemShut {NoStop}%
\bibitem [{\citenamefont {Keldysh}(1964)}]{Keldysh_1965}%
  \BibitemOpen
  \bibfield  {author} {\bibinfo {author} {\bibfnamefont {L.~V.}\ \bibnamefont
  {Keldysh}},\ }\bibfield  {title} {\bibinfo {title} {Ionization in the field
  of a strong electromagnetic wave},\ }\href@noop {} {\bibfield  {journal}
  {\bibinfo  {journal} {Zh. Eksp. Teor. Fiz.}\ }\textbf {\bibinfo {volume}
  {47}},\ \bibinfo {pages} {1945} (\bibinfo {year} {1964})}\BibitemShut
  {NoStop}%
\bibitem [{\citenamefont {Ivanov}\ and\ \citenamefont
  {Kheifets}(2014)}]{Ivanov_2014}%
  \BibitemOpen
  \bibfield  {author} {\bibinfo {author} {\bibfnamefont {I.~A.}\ \bibnamefont
  {Ivanov}}\ and\ \bibinfo {author} {\bibfnamefont {A.~S.}\ \bibnamefont
  {Kheifets}},\ }\bibfield  {title} {\bibinfo {title} {Strong-field ionization
  of he by elliptically polarized light in attoclock configuration},\
  }\href@noop {} {\bibfield  {journal} {\bibinfo  {journal} {Phys. Rev. A}\
  }\textbf {\bibinfo {volume} {89}},\ \bibinfo {pages} {021402} (\bibinfo
  {year} {2014})}\BibitemShut {NoStop}%
\bibitem [{\citenamefont {Muller}(1999)}]{Muller_1999}%
  \BibitemOpen
  \bibfield  {author} {\bibinfo {author} {\bibfnamefont {H.~G.}\ \bibnamefont
  {Muller}},\ }\bibfield  {title} {\bibinfo {title} {An efficient scheme for
  the time-dependent schr\"odinger equation in the velocity gauge},\
  }\href@noop {} {\bibfield  {journal} {\bibinfo  {journal} {Laser Phys.}\
  }\textbf {\bibinfo {volume} {9}},\ \bibinfo {pages} {138} (\bibinfo {year}
  {1999})}\BibitemShut {NoStop}%
\bibitem [{\citenamefont {Bauer}\ and\ \citenamefont
  {Koval}(2006)}]{Bauer_2006}%
  \BibitemOpen
  \bibfield  {author} {\bibinfo {author} {\bibfnamefont {D.}~\bibnamefont
  {Bauer}}\ and\ \bibinfo {author} {\bibfnamefont {P.}~\bibnamefont {Koval}},\
  }\bibfield  {title} {\bibinfo {title} {Qprop: A schr\"odinger-solver for
  intense laser-atom interaction},\ }\href@noop {} {\bibfield  {journal}
  {\bibinfo  {journal} {Computer Phys. Comm.}\ }\textbf {\bibinfo {volume}
  {174}},\ \bibinfo {pages} {396} (\bibinfo {year} {2006})}\BibitemShut
  {NoStop}%
\bibitem [{\citenamefont {Patchkovskii}\ and\ \citenamefont
  {Muller}(2016)}]{Patchkovskii_2016}%
  \BibitemOpen
  \bibfield  {author} {\bibinfo {author} {\bibfnamefont {S.}~\bibnamefont
  {Patchkovskii}}\ and\ \bibinfo {author} {\bibfnamefont {H.~G.}\ \bibnamefont
  {Muller}},\ }\bibfield  {title} {\bibinfo {title} {Simple, accurate, and
  efficient implementation of 1-electron atomic time-dependent schrödinger
  equation in spherical coordinates},\ }\href@noop {} {\bibfield  {journal}
  {\bibinfo  {journal} {Comp. Phys. Commun.}\ }\textbf {\bibinfo {volume}
  {199}},\ \bibinfo {pages} {153 } (\bibinfo {year} {2016})}\BibitemShut
  {NoStop}%
\bibitem [{\citenamefont {Majety}\ and\ \citenamefont
  {Scrinzi}(2017)}]{Majety_2017}%
  \BibitemOpen
  \bibfield  {author} {\bibinfo {author} {\bibfnamefont {V.~P.}\ \bibnamefont
  {Majety}}\ and\ \bibinfo {author} {\bibfnamefont {A.}~\bibnamefont
  {Scrinzi}},\ }\bibfield  {title} {\bibinfo {title} {Absence of electron
  correlation effects in the helium attoclock setting},\ }\href@noop {}
  {\bibfield  {journal} {\bibinfo  {journal} {J. Mod. Optics}\ }\textbf
  {\bibinfo {volume} {64}},\ \bibinfo {pages} {1026} (\bibinfo {year}
  {2017})}\BibitemShut {NoStop}%
\bibitem [{\citenamefont {Popruzhenko}\ \emph {et~al.}(2008)\citenamefont
  {Popruzhenko}, \citenamefont {Paulus},\ and\ \citenamefont
  {Bauer}}]{Popruzhenko_2008a}%
  \BibitemOpen
  \bibfield  {author} {\bibinfo {author} {\bibfnamefont {S.~V.}\ \bibnamefont
  {Popruzhenko}}, \bibinfo {author} {\bibfnamefont {G.~G.}\ \bibnamefont
  {Paulus}},\ and\ \bibinfo {author} {\bibfnamefont {D.}~\bibnamefont
  {Bauer}},\ }\bibfield  {title} {\bibinfo {title} {Coulomb-corrected quantum
  trajectories in strong-field ionization},\ }\href@noop {} {\bibfield
  {journal} {\bibinfo  {journal} {Phys. Rev. A}\ }\textbf {\bibinfo {volume}
  {77}},\ \bibinfo {pages} {053409} (\bibinfo {year} {2008})}\BibitemShut
  {NoStop}%
\bibitem [{\citenamefont {Popruzhenko}\ and\ \citenamefont
  {Bauer}(2008)}]{Popruzhenko_2008b}%
  \BibitemOpen
  \bibfield  {author} {\bibinfo {author} {\bibfnamefont {S.~V.}\ \bibnamefont
  {Popruzhenko}}\ and\ \bibinfo {author} {\bibfnamefont {D.}~\bibnamefont
  {Bauer}},\ }\href@noop {} {\bibfield  {journal} {\bibinfo  {journal} {J. Mod.
  Opt.}\ }\textbf {\bibinfo {volume} {55}},\ \bibinfo {pages} {2573} (\bibinfo
  {year} {2008})}\BibitemShut {NoStop}%
\bibitem [{\citenamefont {Torlina}\ and\ \citenamefont
  {Smirnova}(2012)}]{Torlina_2012}%
  \BibitemOpen
  \bibfield  {author} {\bibinfo {author} {\bibfnamefont {L.}~\bibnamefont
  {Torlina}}\ and\ \bibinfo {author} {\bibfnamefont {O.}~\bibnamefont
  {Smirnova}},\ }\bibfield  {title} {\bibinfo {title} {Time-dependent
  analytical $r$-matrix approach for strong-field dynamics. i. one-electron
  systems},\ }\href@noop {} {\bibfield  {journal} {\bibinfo  {journal} {Phys.
  Rev. A}\ }\textbf {\bibinfo {volume} {86}},\ \bibinfo {pages} {043408}
  (\bibinfo {year} {2012})}\BibitemShut {NoStop}%
\bibitem [{\citenamefont {Kaushal}\ and\ \citenamefont
  {Smirnova}(2013)}]{Kaushal_2013}%
  \BibitemOpen
  \bibfield  {author} {\bibinfo {author} {\bibfnamefont {J.}~\bibnamefont
  {Kaushal}}\ and\ \bibinfo {author} {\bibfnamefont {O.}~\bibnamefont
  {Smirnova}},\ }\bibfield  {title} {\bibinfo {title} {Nonadiabatic coulomb
  effects in strong-field ionization in circularly polarized laser fields},\
  }\href@noop {} {\bibfield  {journal} {\bibinfo  {journal} {Phys. Rev. A}\
  }\textbf {\bibinfo {volume} {88}},\ \bibinfo {pages} {013421} (\bibinfo
  {year} {2013})}\BibitemShut {NoStop}%
\bibitem [{\citenamefont {Nubbemeyer}\ \emph {et~al.}(2008)\citenamefont
  {Nubbemeyer}, \citenamefont {Gorling}, \citenamefont {Saenz}, \citenamefont
  {Eichmann},\ and\ \citenamefont {Sandner}}]{Nubbemeyer_2008}%
  \BibitemOpen
  \bibfield  {author} {\bibinfo {author} {\bibfnamefont {T.}~\bibnamefont
  {Nubbemeyer}}, \bibinfo {author} {\bibfnamefont {K.}~\bibnamefont {Gorling}},
  \bibinfo {author} {\bibfnamefont {A.}~\bibnamefont {Saenz}}, \bibinfo
  {author} {\bibfnamefont {U.}~\bibnamefont {Eichmann}},\ and\ \bibinfo
  {author} {\bibfnamefont {W.}~\bibnamefont {Sandner}},\ }\bibfield  {title}
  {\bibinfo {title} {Strong-field tunneling without ionization},\ }\href@noop
  {} {\bibfield  {journal} {\bibinfo  {journal} {Phys. Rev. Lett.}\ }\textbf
  {\bibinfo {volume} {101}},\ \bibinfo {pages} {233001} (\bibinfo {year}
  {2008})}\BibitemShut {NoStop}%
\bibitem [{\citenamefont {Feuerstein}\ and\ \citenamefont
  {Thumm}(2003)}]{Feuerstein_2003}%
  \BibitemOpen
  \bibfield  {author} {\bibinfo {author} {\bibfnamefont {B.}~\bibnamefont
  {Feuerstein}}\ and\ \bibinfo {author} {\bibfnamefont {U.}~\bibnamefont
  {Thumm}},\ }\bibfield  {title} {\bibinfo {title} {{On the computation of
  momentum distributions within wavepacket propagation calculations}},\
  }\href@noop {} {\bibfield  {journal} {\bibinfo  {journal} {J. Phys. B}\
  }\textbf {\bibinfo {volume} {36}},\ \bibinfo {pages} {707} (\bibinfo {year}
  {2003})}\BibitemShut {NoStop}%
\bibitem [{\citenamefont {Wang}\ \emph {et~al.}(2013)\citenamefont {Wang},
  \citenamefont {Tian},\ and\ \citenamefont {Eberly}}]{Wang_2013}%
  \BibitemOpen
  \bibfield  {author} {\bibinfo {author} {\bibfnamefont {X.}~\bibnamefont
  {Wang}}, \bibinfo {author} {\bibfnamefont {J.}~\bibnamefont {Tian}},\ and\
  \bibinfo {author} {\bibfnamefont {J.~H.}\ \bibnamefont {Eberly}},\ }\bibfield
   {title} {\bibinfo {title} {{Extended Virtual Detector Theory for
  Strong-Field Atomic Ionization}},\ }\href@noop {} {\bibfield  {journal}
  {\bibinfo  {journal} {Phys. Rev. Lett.}\ }\textbf {\bibinfo {volume} {110}},\
  \bibinfo {pages} {243001} (\bibinfo {year} {2013})}\BibitemShut {NoStop}%
\bibitem [{Sup()}]{Supplement}%
  \BibitemOpen
  \href@noop {} {}\bibinfo {howpublished} {See the Supplemental Materials for
  the details.}\BibitemShut {Stop}%
\bibitem [{\citenamefont {Faisal}(1973)}]{Faisal_1973}%
  \BibitemOpen
  \bibfield  {author} {\bibinfo {author} {\bibfnamefont {F.~H.~M.}\
  \bibnamefont {Faisal}},\ }\bibfield  {title} {\bibinfo {title} {Multiple
  absorption of laser photons by atoms},\ }\href@noop {} {\bibfield  {journal}
  {\bibinfo  {journal} {J. Phys. B}\ }\textbf {\bibinfo {volume} {6}},\
  \bibinfo {pages} {L89} (\bibinfo {year} {1973})}\BibitemShut {NoStop}%
\bibitem [{\citenamefont {Reiss}(1980)}]{Reiss_1980}%
  \BibitemOpen
  \bibfield  {author} {\bibinfo {author} {\bibfnamefont {H.~R.}\ \bibnamefont
  {Reiss}},\ }\bibfield  {title} {\bibinfo {title} {Effect of an intesne
  electromagnetic field on a weakly bound system},\ }\href@noop {} {\bibfield
  {journal} {\bibinfo  {journal} {Phys. Rev. A}\ }\textbf {\bibinfo {volume}
  {22}},\ \bibinfo {pages} {1786} (\bibinfo {year} {1980})}\BibitemShut
  {NoStop}%
\bibitem [{\citenamefont {Kroll}\ and\ \citenamefont
  {Watson}(1973)}]{Kroll_1973}%
  \BibitemOpen
  \bibfield  {author} {\bibinfo {author} {\bibfnamefont {N.~M.}\ \bibnamefont
  {Kroll}}\ and\ \bibinfo {author} {\bibfnamefont {K.~M.}\ \bibnamefont
  {Watson}},\ }\bibfield  {title} {\bibinfo {title} {Charged-particle
  scattering in the presence of a strong electromagnetic wave},\ }\href@noop {}
  {\bibfield  {journal} {\bibinfo  {journal} {Phys. Rev. A}\ }\textbf {\bibinfo
  {volume} {8}},\ \bibinfo {pages} {804} (\bibinfo {year} {1973})}\BibitemShut
  {NoStop}%
\bibitem [{\citenamefont {\ifmmode \check{C}\else
  \v{C}\fi{}erki\ifmmode~\acute{c}\else \'{c}\fi{}}\ \emph
  {et~al.}(2009)\citenamefont {\ifmmode \check{C}\else
  \v{C}\fi{}erki\ifmmode~\acute{c}\else \'{c}\fi{}}, \citenamefont
  {Hasovi\ifmmode~\acute{c}\else \'{c}\fi{}}, \citenamefont {Milo\ifmmode
  \check{s}\else \v{s}\fi{}evi\ifmmode~\acute{c}\else \'{c}\fi{}},\ and\
  \citenamefont {Becker}}]{Cerkic_2009}%
  \BibitemOpen
  \bibfield  {author} {\bibinfo {author} {\bibfnamefont {A.}~\bibnamefont
  {\ifmmode \check{C}\else \v{C}\fi{}erki\ifmmode~\acute{c}\else \'{c}\fi{}}},
  \bibinfo {author} {\bibfnamefont {E.}~\bibnamefont
  {Hasovi\ifmmode~\acute{c}\else \'{c}\fi{}}}, \bibinfo {author} {\bibfnamefont
  {D.~B.}\ \bibnamefont {Milo\ifmmode \check{s}\else
  \v{s}\fi{}evi\ifmmode~\acute{c}\else \'{c}\fi{}}},\ and\ \bibinfo {author}
  {\bibfnamefont {W.}~\bibnamefont {Becker}},\ }\bibfield  {title} {\bibinfo
  {title} {High-order above-threshold ionization beyond the first-order born
  approximation},\ }\href@noop {} {\bibfield  {journal} {\bibinfo  {journal}
  {Phys. Rev. A}\ }\textbf {\bibinfo {volume} {79}},\ \bibinfo {pages} {033413}
  (\bibinfo {year} {2009})}\BibitemShut {NoStop}%
\bibitem [{\citenamefont {Milo\ifmmode \check{s}\else
  \v{s}\fi{}evi\ifmmode~\acute{c}\else \'{c}\fi{}}(2014)}]{Milosevic_2014a}%
  \BibitemOpen
  \bibfield  {author} {\bibinfo {author} {\bibfnamefont {D.~B.}\ \bibnamefont
  {Milo\ifmmode \check{s}\else \v{s}\fi{}evi\ifmmode~\acute{c}\else
  \'{c}\fi{}}},\ }\bibfield  {title} {\bibinfo {title} {Low-frequency
  approximation for above-threshold ionization by a laser pulse: Low-energy
  forward rescattering},\ }\href@noop {} {\bibfield  {journal} {\bibinfo
  {journal} {Phys. Rev. A}\ }\textbf {\bibinfo {volume} {90}},\ \bibinfo
  {pages} {063423} (\bibinfo {year} {2014})}\BibitemShut {NoStop}%
\bibitem [{\citenamefont {Becker}\ \emph {et~al.}(2002)\citenamefont {Becker},
  \citenamefont {Grasbon}, \citenamefont {Kopold}, \citenamefont
  {Milo\u{s}evi\'c}, \citenamefont {Paulus},\ and\ \citenamefont
  {Walther}}]{Becker_2002}%
  \BibitemOpen
  \bibfield  {author} {\bibinfo {author} {\bibfnamefont {W.}~\bibnamefont
  {Becker}}, \bibinfo {author} {\bibfnamefont {F.}~\bibnamefont {Grasbon}},
  \bibinfo {author} {\bibfnamefont {R.}~\bibnamefont {Kopold}}, \bibinfo
  {author} {\bibfnamefont {D.~B.}\ \bibnamefont {Milo\u{s}evi\'c}}, \bibinfo
  {author} {\bibfnamefont {G.~G.}\ \bibnamefont {Paulus}},\ and\ \bibinfo
  {author} {\bibfnamefont {H.}~\bibnamefont {Walther}},\ }\bibfield  {title}
  {\bibinfo {title} {Above-threshold ionization: from classical features to
  quantum effects},\ }\href@noop {} {\bibfield  {journal} {\bibinfo  {journal}
  {Adv. Atom. Mol. Opt. Phys.}\ }\textbf {\bibinfo {volume} {48}},\ \bibinfo
  {pages} {35} (\bibinfo {year} {2002})}\BibitemShut {NoStop}%
\bibitem [{\citenamefont {Wolkow}(1935)}]{Volkov_1935}%
  \BibitemOpen
  \bibfield  {author} {\bibinfo {author} {\bibfnamefont {D.~M.}\ \bibnamefont
  {Wolkow}},\ }\bibfield  {title} {\bibinfo {title} {{\"Uber eine Klasse von
  L\"osungen der Diracschen Gleichung}},\ }\href@noop {} {\bibfield  {journal}
  {\bibinfo  {journal} {Z. Phys.}\ }\textbf {\bibinfo {volume} {94}},\ \bibinfo
  {pages} {250} (\bibinfo {year} {1935})}\BibitemShut {NoStop}%
\bibitem [{\citenamefont {Augst}\ \emph {et~al.}(1989)\citenamefont {Augst},
  \citenamefont {Strickland}, \citenamefont {Meyerhofer}, \citenamefont
  {Chin},\ and\ \citenamefont {Eberly}}]{Augst_1989}%
  \BibitemOpen
  \bibfield  {author} {\bibinfo {author} {\bibfnamefont {S.}~\bibnamefont
  {Augst}}, \bibinfo {author} {\bibfnamefont {D.}~\bibnamefont {Strickland}},
  \bibinfo {author} {\bibfnamefont {D.~D.}\ \bibnamefont {Meyerhofer}},
  \bibinfo {author} {\bibfnamefont {S.~L.}\ \bibnamefont {Chin}},\ and\
  \bibinfo {author} {\bibfnamefont {J.~H.}\ \bibnamefont {Eberly}},\ }\bibfield
   {title} {\bibinfo {title} {Tunneling ionization of noble gases in a
  high-intensity laser field},\ }\href@noop {} {\bibfield  {journal} {\bibinfo
  {journal} {Phys. Rev. Lett.}\ }\textbf {\bibinfo {volume} {63}},\ \bibinfo
  {pages} {2212} (\bibinfo {year} {1989})}\BibitemShut {NoStop}%
\bibitem [{\citenamefont {Yakaboylu}\ \emph {et~al.}(2014)\citenamefont
  {Yakaboylu}, \citenamefont {Klaiber},\ and\ \citenamefont
  {Hatsagortsyan}}]{Yakaboylu_2014b}%
  \BibitemOpen
  \bibfield  {author} {\bibinfo {author} {\bibfnamefont {E.}~\bibnamefont
  {Yakaboylu}}, \bibinfo {author} {\bibfnamefont {M.}~\bibnamefont {Klaiber}},\
  and\ \bibinfo {author} {\bibfnamefont {K.~Z.}\ \bibnamefont
  {Hatsagortsyan}},\ }\bibfield  {title} {\bibinfo {title} {Wigner time delay
  for tunneling ionization via the electron propagator},\ }\href@noop {}
  {\bibfield  {journal} {\bibinfo  {journal} {Phys. Rev. A}\ }\textbf {\bibinfo
  {volume} {90}},\ \bibinfo {pages} {012116} (\bibinfo {year}
  {2014})}\BibitemShut {NoStop}%
\bibitem [{\citenamefont {Peres}(1980)}]{Peres_1980}%
  \BibitemOpen
  \bibfield  {author} {\bibinfo {author} {\bibfnamefont {A.}~\bibnamefont
  {Peres}},\ }\bibfield  {title} {\bibinfo {title} {Measurement of time by
  quantum clocks},\ }\href@noop {} {\bibfield  {journal} {\bibinfo  {journal}
  {Am. J. Phys.}\ }\textbf {\bibinfo {volume} {48}},\ \bibinfo {pages} {552}
  (\bibinfo {year} {1980})}\BibitemShut {NoStop}%
\bibitem [{\citenamefont {Klaiber}\ \emph {et~al.}(2013)\citenamefont
  {Klaiber}, \citenamefont {Yakaboylu}, \citenamefont {Bauke}, \citenamefont
  {Hatsagortsyan},\ and\ \citenamefont {Keitel}}]{Klaiber_2013c}%
  \BibitemOpen
  \bibfield  {author} {\bibinfo {author} {\bibfnamefont {M.}~\bibnamefont
  {Klaiber}}, \bibinfo {author} {\bibfnamefont {E.}~\bibnamefont {Yakaboylu}},
  \bibinfo {author} {\bibfnamefont {H.}~\bibnamefont {Bauke}}, \bibinfo
  {author} {\bibfnamefont {K.~Z.}\ \bibnamefont {Hatsagortsyan}},\ and\
  \bibinfo {author} {\bibfnamefont {C.~H.}\ \bibnamefont {Keitel}},\ }\bibfield
   {title} {\bibinfo {title} {Under-the-barrier dynamics in laser-induced
  relativistic tunneling},\ }\href@noop {} {\bibfield  {journal} {\bibinfo
  {journal} {Phys. Rev. Lett.}\ }\textbf {\bibinfo {volume} {110}},\ \bibinfo
  {pages} {153004} (\bibinfo {year} {2013})}\BibitemShut {NoStop}%
\bibitem [{\citenamefont {Can\'ario}\ \emph {et~al.}(2021)\citenamefont
  {Can\'ario}, \citenamefont {Klaiber}, \citenamefont {Hatsagortsyan},\ and\
  \citenamefont {Keitel}}]{Canario_2021}%
  \BibitemOpen
  \bibfield  {author} {\bibinfo {author} {\bibfnamefont {D.~B.}\ \bibnamefont
  {Can\'ario}}, \bibinfo {author} {\bibfnamefont {M.}~\bibnamefont {Klaiber}},
  \bibinfo {author} {\bibfnamefont {K.~Z.}\ \bibnamefont {Hatsagortsyan}},\
  and\ \bibinfo {author} {\bibfnamefont {C.~H.}\ \bibnamefont {Keitel}},\
  }\bibfield  {title} {\bibinfo {title} {Role of reflections in the generation
  of a time delay in strong-field ionization},\ }\href
  {https://doi.org/10.1103/PhysRevA.104.033103} {\bibfield  {journal} {\bibinfo
   {journal} {Phys. Rev. A}\ }\textbf {\bibinfo {volume} {104}},\ \bibinfo
  {pages} {033103} (\bibinfo {year} {2021})}\BibitemShut {NoStop}%
\end{thebibliography}%

\end{document}